\documentclass[%
reprint,
amsmath,amssymb,
prb,
]{revtex4-1}

\usepackage{graphicx}
\usepackage{dcolumn}
\usepackage{bm}
\usepackage{xcolor}
\usepackage{upgreek}

\begin{document}

\preprint{APS/123-QED}

\title{Modelling the transfer function of two-dimensional SQUID and SQIF arrays with thermal noise}

\author{Marc A. Gal\'i Labarias}
\email{marc.galilabarias@csiro.au}
\author{Karl-H. M\"uller}
\author{Emma E. Mitchell}%
\affiliation{%
 CSIRO Manufacturing, Lindfield, NSW, Australia.
}%




\date{\today}

\begin{abstract}
We present a theoretical model for 2D SQUID and SQIF arrays with over-damped Josephson junctions for uniform bias current injection at 77 K. Our simulations demonstrate the importance of including Johnson thermal noise and reveal that the mutual inductive coupling between SQUID loops is of minor importance. Our numerical results establish the validity of a simple scaling behaviour between the voltages of 1D and 2D SQUID arrays and show that the same scaling behaviour applies to the maximum transfer functions. The maximum transfer function of a 2D SQUID array can be further optimised by applying the optimal bias current which depends on the SQUID loop self-inductance and the junction critical current. Our investigation further reveals that a scaling behaviour exits between the maximum transfer function of a 2D SQUID array and that of a single dc-SQUID. Finally, we investigate the voltage response of 1D and 2D SQIF arrays and illustrate the effects of adding spreads in the heights and widths of SQUID loops.
\end{abstract}

\maketitle

\section{\label{sec:Intro}Introduction}

Superconducting quantum interference devices (SQUIDs) have been extensively investigated for their very high magnetic field sensitivity both experimentally \cite{} and theoretically \cite{Clarke2004, Clem2005} . 
SQUIDs are routinely fabricated using both low-temperature superconducting (LTS) and high-temperature superconducting (HTS) thin films.
The lower operating temperature provides LTS SQUIDs with better noise performance at the expense of more complex cryogenic conditions, compared to HTS SQUIDs.
For example, HTS SQUIDs have found applications in geophysical exploration \cite{Foley1999a}, due to less stringent cryogenic constraints resulting in SQUID systems with reduced size, weight and power.
\citet{Voss1981} studied the effect of thermal noise on the I-V characteristics of shunted Josephson junctions (JJs), and \citet{Tesche1977} and \citet{Enpuku1993} extensively investigated the effect of thermal noise on the performance of dc-SQUIDs.
These studies showed a significant decrease in the voltage modulation depth and transfer function due to thermal noise at high temperatures (77 K).

The interest in SQUID arrays appeared in order to improve the sensitivity and robustness of dc-SQUIDs.
\citet{Miller1991} theoretically studied one-dimensional (1D) SQUID arrays, also called superconducting quantum interference grating (SQUIG) as an analogy to optical interference gratings. 
In this work the authors considered self- and mutual-inductance effects of the SQUIDs connected in parallel, as well as different screening parameters. 
Their results predicted a better magnetic field resolution for 1D parallel SQUID arrays compared to a single dc-SQUID. 
Despite this prediction, \citet{Gerdemann1995} experimentally measured small 1D parallel arrays of HTS SQUIDs, showing a decrease of the voltage modulation, which was attributed to bias current-induced magnetic flux. 
Likewise, \citet{Mitchell2019} experimentally found a degradation in the voltage modulation with the number N of junctions in parallel for $N>11$ and modelling predicted either a plateau or a decrease of the transfer function with $N$ depending on the bias current lead configuration.
Early measurements of 1D arrays of M SQUIDs connected in series \cite{Foglietti1993, Krey1999} showed voltage modulation improvement and white noise reduction with increasing M.

One-dimensional parallel SQUID and SQIF arrays have been previously theoretically studied at $T=0K$ \cite{Berggren2012, Berggren2015, Mitchell2019} and experimentally investigated at high temperatures ($T=77K$) \cite{Mitchell2019}. 
Recently, \citet{Muller2021} introduced a theoretical model for 1D parallel HTS SQUID arrays that includes thermal noise and fluxoid focusing. This model showed excellent agreement with experimental results at 77K.

Superconducting quantum interference filters (SQIF) were theoretically proposed by \citet{Oppenlander2000} and experimentally reported at high temperatures by \citet{Caputo2005}. 
The SQUIDs making up these arrays have different loop areas creating a destructive voltage interference for magnetic fluxes away from zero external magnetic flux. Therefore SQIF arrays are ideal to perform absolute magnetic field measurements, since they present a unique dip.
Obtaining a SQIF-like response using SQUIDs in series was theoretically studied by \citet{Haussler2001} and then experimentally demonstrated by \citet{Oppenlander2002}. 
Alternatively, \citet{Longhini2011} varied the distance between the non-locally coupled SQUIDs. In doing so, the magnetic coupling between SQUIDs differs which breaks the periodicity of the voltage with the applied magnetic field.
Interest has grown in the performance of two-dimensional (2D) SQIF arrays which are predicted to have improved sensitivity, dynamic range, bandwidth and linearity compared with single SQUIDs \cite{Schultze2006}.
Kornev \emph{et al.}\cite{Kornev2009b, Kornev2011} studied 1D parallel SQIF arrays connected in series and analysed the SQUID coupling in the array. They also showed the linear increase of the voltage modulation with the number of SQIFs in series and the conservation of linearity. 

Large 2D SQUID arrays operating at high-temperatures have been experimentally measured and studied \cite{Mitchell2016, Taylor2016}. 
Two-dimensional SQUID arrays models without considering thermal noise have been previously investigated by \citet{Cybart2012}, \citet{Dalichaouch2014} and \citet{Taylor2016}, but these models become inaccurate at 77 $K$ where the thermal noise strength is large.

The goal of this work is to introduce a model that accurately calculates the response of 2D SQUID and SQIF arrays operating at 77 K by including the thermal noise from the junction resistors. 
Our theoretical model assumes overdamped junctions (RSJ model) and includes the magnetic flux coupling due to all the currents flowing in the array as well as the  magnetic flux created by the bias leads. 
The conservation of currents at every vertex of the array is taken into account. 
In this work we will demonstrate that including thermal noise is crucial to obtain the correct array response for devices operating at high-temperatures (77 K), where the thermal noise plays an important role.
We will use this model to compare the time-averaged voltage and the maximum transfer function of 1D and 2D SQUID arrays at $T=77$ K for different bias currents.
We will show that the mutual inductive coupling between SQUID loops is of minor importance to calculate the maximum transfer function and the overall voltage response. Most importantly, we will demonstrate that the voltage response of 1D and 2D SQUID arrays are approximately proportional. And that the maximum transfer function of a 2D SQUID array can be directly related to the maximum transfer function of a dc-SQUID.
Furthermore, we will compare 2D SQIF arrays with different SQUID area distributions.

Our paper is structured as follows. In Sec. \ref{sec:The} we introduce the mathematical framework of our model for 2D SQUID and SQIF arrays. We derive the system of coupled differential equations for the phase differences of the overdamped JJ's of the array where we include the effects of the mutual inductive coupling between the SQUID loops and the Johnson thermal noise from the JJ resistors.
In Sec. \ref{sec:Sim} we present our simulation results. 
In (A) we emphasise the importance of thermal noise and in (B) we study the voltage response of 2D SQUID arrays and the effect of the mutual inductive coupling between SQUID loops. 
In (C) we investigate the dependence of the maximum transfer function on the device bias current and array size, while in (D) we reveal its dependence on the JJ critical current and the SQUID loop self-inductance.
In (E) we show how the maximum transfer function of a 2D SQUID array is related to that of a dc-SQUID.
In (F) we discuss the voltage modulation depth of 2D SQUID arrays.
And in (G) we present our findings for 2D SQIF arrays.
Finally, in Sec. \ref{sec:Con} we give a summary of our work.

\section{Mathematical Model}\label{sec:The}

\begin{figure}
    \centering
    \includegraphics[scale=0.33]{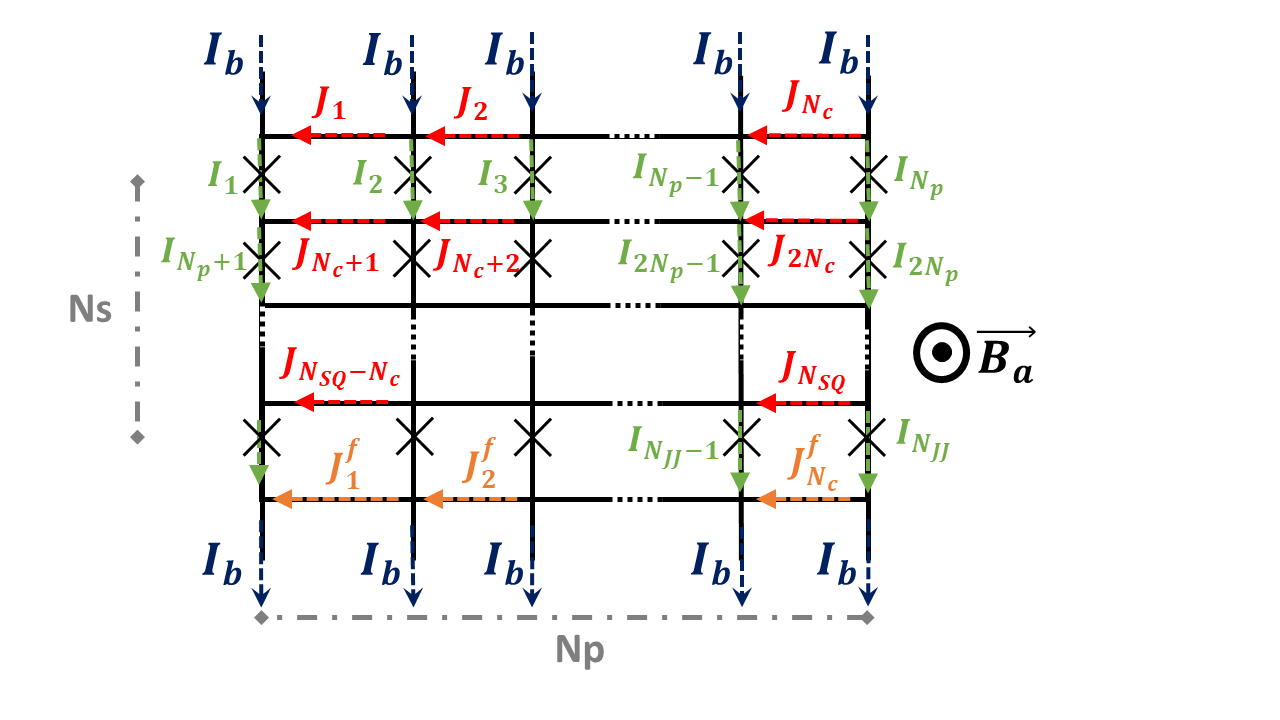}
    \caption{Diagram of the different currents used in our model for SQUID and SQIF arrays. Arrows indicate the current direction used for Kirchhoff's Law analysis. In dark blue we show the time-independent bias currents $I_b$. Green arrows represent the vertical currents $I_k$, which have the same direction as the bias currents. Red depicts the horizontal currents $J_k$, and orange arrows represent the horizontal currents $J^f_k$ flowing along the bottom part of the SQUIDs in the last row. Black crosses indicate JJs. The applied magnetic field $\vec{B}_a$ points upwards perpendicular to the array as indicated. $N_p$ is the number of JJs connected in parallel per row, while $N_s$ is number of rows. Finally $N_c=N_p-1$ is the number of SQUIDs in parallel (number of columns).}
    \label{fig:diag}
\end{figure}

In Fig. \ref{fig:diag} we show a schematic diagram of the 2D SQUID and SQIF arrays under study, which consist of SQUIDs connected by sharing JJs along their sides. The loop areas are identical for SQUID arrays while they differ for SQIF arrays.
We use the notation $(N_s, N_p)$-array which is an array with $N_s$ JJs in series and $N_p$ JJs in parallel.
The $(1,2)$-array is the common dc-SQUID, a 1D parallel array is a $(1, N_p)$-array and a serial dc-SQUID array is a $(N_s, 2)$-array.
These arrays have $N_c=N_p-1$ number of SQUIDs in each row, a total number of SQUIDs of $N_{SQ}=N_c \times N_s$ and a total number of JJs of $N_{JJ}=N_p \times N_s$.
In this study we assume a grid-like structure, where the heights of the SQUIDs in the same row are equal. The same holds for the widths of the SQUIDs in the same column.
In our arrays the JJs are located only in the vertical tracks, which is very different to the so-called JJ-arrays where JJs are also present along the horizontal tracks \cite{Newrock2000}. 

For this study we are assuming uniformly biased arrays, \emph{i.e.} there are $N_p$ entering and exiting bias leads which carry equal currents $I_b$ (Fig. \ref{fig:diag}).

{}

\begin{figure}
    \centering
    \includegraphics[scale=0.3]{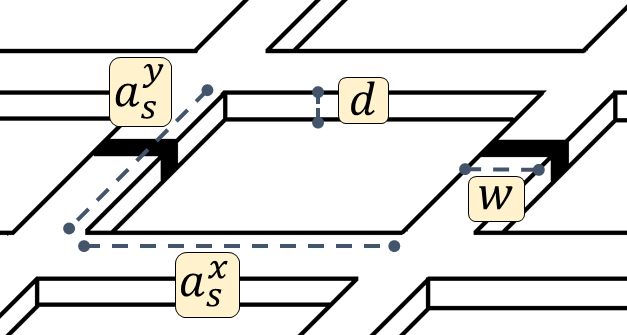}
    \caption{Sketch of the $s^{\text{th}}$ SQUID in the array. $a^x_s$ and $a^y_s$ are the width and height of the $s^{\text{th}}$ SQUID loop, $w$ is the track width and $d$ the film thickness. The JJs are represented by thick black cuts in the vertical loop tracks.}
    \label{fig:SQUID-geo}
\end{figure}

As the SQUID loops are rectangular it is convenient to define vertical currents $I_k$ and horizontal currents $J_k$ (Fig. \ref{fig:diag}).
Assuming identical overdamped JJs (RSJ model) one derives the current-phase equation
\begin{align}
    I_k(t) + I^{n}_k(t) &= I_c \sin \varphi_k(t) + \frac{\Phi_0}{2\pi R}\frac{d \varphi_k(t)}{dt},
    \label{eq:current-phase-eq}
\end{align}{}
where $t$ is the time, $I_c$ and $R$ are the critical current and normal resistance of the JJs and $\Phi_0$ the flux quantum. $\varphi_k(t)$ is the time-dependent gauge-invariant phase difference across the $k^{\text{th}}$ junction and $I^n_k(t)$ is the noise current created by the Johnson thermal noise at the $k^{\text{th}}$ junction. 

\subsection{Kirchhoff's Law}
At each vertex (crossing point between tracks) one can apply Kirchhoff's law which gives
\begin{align}
    & I_k = J_k - J_{k-1} + I_b , \label{eq:cons-current-top}\\
    & I_{(n-1)N_p+k} = J_{(n-1)N_c+k} - J_{(n-1)N_c + (k-1)} \nonumber \\
    & \qquad  + I_{(n-2)N_p+k} , \label{eq:cons-current} 
\end{align}{}
with $1 < k < N_p$ and $1 < n \leq N_s$. Here Eq. (\ref{eq:cons-current-top}) corresponds to any inside vertex of the top row, and Eq. (\ref{eq:cons-current}) defines any inside vertex of any other row (\emph{i.e.} $ 1 < n \leq N_s$).
For the first vertex of each row one finds
\begin{align}
    &I_1 = J_1  + I_b, \\
    &I_{(n-1)N_p + 1} = J_{(n-1)N_c + 1}  + I_{(n-2)N_p + 1} .
\end{align}{}

Equivalently, for the last vertex of each row one obtains
\begin{align}
    & I_{N_p} =  - J_{N_c} + I_b, \\
    & I_{nN_p} =  - J_{nN_c} + I_{(n-1)N_p} .
    \label{eq:cons-current-right}
\end{align}{}

Using matrix notation we can combine Eqs. (\ref{eq:cons-current-top})-(\ref{eq:cons-current-right}) and obtain
\begin{align}
    & \hat{K}_I \vec{I} = \hat{K}_J \vec{J} + \vec{I}_b, \\
    & \vec{I} = \hat{K}_I^{-1} \left( \hat{K}_J \vec{J} + \vec{I}_b \right),
    \label{eq:I_kirch}
\end{align}{}
where $\hat{K}_I$ is a square matrix with dimensions $[N_{JJ} \times N_{JJ}]$, and $\hat{K}_J$ is a matrix with dimensions $[N_{JJ} \times N_{SQ}]$. The elements of these two matrices are defined as
\begin{align}
    (\hat{K}_I )_{ij} &= \delta_{i,j} - \delta_{i - N_p, j} , \\
    ( \hat{K}_J )_{ij} &=  \delta_{i,j} - \delta_{i-1,j},
\end{align}
where $\delta_{i,j}$ is the Kronecker delta.

The current vectors are defined as
\begin{align}
    \vec{I} &= (I_1, I_2, \dots, I_{N_{JJs}})^T, \\
    \vec{J} &= (J_1, J_2, \dots, J_{N_{SQ}})^T, \\
    \vec{I}_b &= (I_b, I_b,\dots, I_b, 0, \dots, 0)^T, \\
    \vec{I}_f &= (I_b, I_b,\dots, I_b)^T,
    \label{eq:current-vectors}
\end{align}{}
where the superscript $T$ means transposition. Note that $\vec{J}$ does not contain the $J_k$ currents of the bottom horizontal tracks. The bias current vector $\vec{I}_b$ has dimension $[N_{JJ}, 1]$ with the first $N_p$ components equal to $I_b$ and the rest being zero. And the bias current vector for the leads exiting from the bottom of the array $\vec{I}_f$ has dimension $[N_p, 1]$.

\subsection{Geometric and kinetic inductance}

Using the second Ginzburg-Landau equation \cite{Tinkham2004}, we can find a relationship between phases, fluxes and currents of the array. As we only have JJs at the vertical sides of each loop, we can choose a closed path around each loop which will connect the total magnetic flux threading each SQUID with the phases of its two junctions and one obtains
\begin{align}
    \frac{\Phi_0}{2\pi}\left( \varphi_{k+1} - \varphi_k \right) &= \Phi^a_s + \Phi_s^L + \mu_0 \lambda^2 \oint_{\mathcal{C}_s} \vec{j} \cdot \vec{dl},
    \label{eq:gauge-inv0}
\end{align}{}
where $\mu_0$ is the permeability of vacuum, $\lambda$ the London penetration depth \cite{London1935} of the material and $\vec{j}$ is the current density along the closed anti-clockwise path $\mathcal{C}_s$ which encircles the $s^{\text{th}}$ SQUID. 
Here $s=k-(n-1)$ where $k$ is the JJ index and $n$ the row index.
In Eq. (\ref{eq:gauge-inv0}), $\Phi_s^a$ is the applied flux threading the $s^{\text{th}}$ SQUID and $\Phi_s^L$ is the flux threading loop number $s$ generated by all the currents flowing in the array (including the leads). 
The applied magnetic flux is $\Phi^a_s = B_a \cdot a^x_s \cdot a^y_s$, with $B_a$ the perpendicular applied magnetic field (Fig. \ref{fig:diag}), and $a_s^x$ and $a_s^y$ the width and height of the $s^{\text{th}}$ SQUID loop (Fig. \ref{fig:SQUID-geo}).

The flux $\Phi_s^L$ can be expressed in terms of the partial geometric inductances ($L$'s) and currents as
\begin{align}
    \Phi^L_s &= \underbrace{\sum_n^{N_{JJ}} L^v_{sn} I_n + \sum_n^{N_{SQ}}L^H_{sn} J_n + \sum_n^{Nc} L^{hf}_{sn} J^f_n}_{\text{magnetic flux created by the array tracks}} \nonumber \\
            &  \underbrace{+ I_b \sum_n^{N_p} L^b_{sn}}_{\text{magnetic flux created by the bias leads}},
            \label{eq:induced_flux}
\end{align}{}
$L^j_{sn}$ are the partial inductance terms, where the first subscript, $s$, defines the SQUID loop where the magnetic flux is induced, and the second subscript defines the current creating that flux. The superscript indicates the different superconductor tracks, \emph{i.e.} vertical ($v$), horizontal ($H$), bottom horizontal tracks of the array ($hf$), and bias leads ($b$) where $L^b_{sn} = L^{in}_{sn} + L^{out}_{sn}$ with $L^{in}_{sn}$ and $L^{out}_{sn}$ the partial inductances of the top and bottom bias leads.

Because in our case the Pearl penetration depth \cite{Pearl1964} $\Lambda = \lambda^2/d$ ($d$ is the film thickness) satisfies $w/2 \lesssim \Lambda$, the current density $\vec{j}$ is approximately homogeneous across tracks.
Therefore the last term in Eq. (\ref{eq:gauge-inv0}) becomes
\begin{align}
    \mu_0 \lambda^2 \oint_{\mathcal{C}_s} \vec{j} \cdot \vec{dl} =  \frac{\mu_0 \Lambda}{w} \left( a^y_s\left[ I_k - I_{k+1} \right] + a^x_s \left[ J_s - J_{s+N_c} \right] \right).
\end{align}

The terms $\frac{\mu_0 \Lambda}{w}a_s^x$ and $\frac{\mu_0 \Lambda}{w}a_s^y$ are the partial kinetic inductances of the $s^{\text{th}}$ SQUID loop. 
In order to simplify notation, from this point onward, we incorporate the partial kinetic inductance terms into the partial geometric self-inductances in Eq. (\ref{eq:induced_flux}). 

Writing Eqs. (\ref{eq:gauge-inv0}) and (\ref{eq:induced_flux}) in matrix notation, one derives
\begin{align}
    \frac{\Phi_0}{2\pi}\hat{D}\vec{\varphi} =& \vec{\Phi}_a + \hat{L}_v \vec{I} + \hat{L}_H \vec{J} + \hat{L}_{hf} \vec{J}_{f} \nonumber \\
    & + \hat{L}_{b} \vec{I}_{b},
    \label{eq:gauge-inv_M}
\end{align}{}
with  $\vec{\varphi} = (\varphi_1, \varphi_2, \dots, \varphi_{N_{JJs}})^T$ and $\vec{\Phi}_a=(\Phi^a_1, \Phi^a_2, \dots, \Phi^a_{N_{SQ}})^T$. 

$\hat{D}$ is a matrix of dimensions $[N_{SQ} \times N_{JJ}]$ defined as
\begin{align}
    \hat{D}_{ij} = \delta_{i, j-1} - \delta_{i,j}.
    \label{eq:D}
\end{align}

\subsection{Phase-difference dynamics of the array}

To derive a system of coupled differential equations for the phase-differences $\varphi_k$ that describes the array dynamics, we need to express Eq. (\ref{eq:current-phase-eq}) in terms of the time-dependent phase-differences $\varphi_k(t)$ and time-independent quantities. To achieve this, we start by writing Eq. (\ref{eq:gauge-inv_M}) in terms of the horizontal currents $J_k$.

We note that the current-vector $\vec{J}_{f}$ (Fig. \ref{fig:diag}) can be expressed in terms of $\vec{I}$ and $\vec{I}_f$ as
\begin{align}
    \vec{J}_f = \hat{N}_f \vec{I}_f - \hat{N}_I \vec{I},
    \label{eq:Jf}
\end{align}{}
where the matrices $\hat{N}_f$ and $\hat{N}_I$ ensure conservation of current at the vertices at the bottom part of the array. The dimensions of $\hat{N}_f$ and $\hat{N}_I$ are $[N_c \times N_p]$ and $[N_c \times N_{JJ}]$ respectively, and these matrices are defined as
\begin{align}
    \left(\hat{N}_f\right)_{ij} &= \sum_{k=1}^{i} \delta_{k, j},
    \\
    \left(\hat{N}_I\right)_{ij} &= \sum_{k=1}^{i} \delta_{N_{JJs} - N_p + k, j}.
    \label{eq:N}
\end{align}

Using Eqns (\ref{eq:D})-(\ref{eq:N}), we rewrite Eq. (\ref{eq:gauge-inv_M}) as
\begin{align}
    \frac{\Phi_0}{2\pi}\hat{D}\vec{\varphi}  =& \vec{\Phi}_a + \hat{L}_V \vec{I} + \hat{L}_H \vec{J}  
    + \hat{L}_{b} \vec{I}_{b}  +  \hat{L}_{hf} \hat{N}_f \vec{I}_f,
    \label{eq:induced_flux_M2}
\end{align}{}
where we have defined $\hat{L}_V = \left( \hat{L}_v - \hat{L}_{hf} \hat{N}_I \right)$.

Finally using conservation of currents, Eq. (\ref{eq:I_kirch}), we can express the phase-current equation, Eq. (\ref{eq:induced_flux_M2}), only in terms of the $\vec{J}$ currents, which gives
\begin{align}
     \frac{\Phi_0}{2\pi}\hat{D}\vec{\varphi} = \vec{\Phi}_{nf} + \hat{L} \vec{J} ,
    \label{eq:phase-current2}
\end{align}{}
where $\hat{L}= \left( \hat{L}_V \hat{K}  + \hat{L}_H \right) $ with $\hat{K}=\hat{K}_I^{-1}\hat{K}_J$, and $\vec{\Phi}_{nf} = \vec{\Phi}_a + \left(\hat{L}_V\hat{K}_I^{-1} + \hat{L}_{b} \right) \vec{I}_{b} + \hat{L}_{hf} \hat{N}_f \vec{I}_{f}$ are time-independent vectors.

Next step is to express Eq. (\ref{eq:current-phase-eq}) in matrix form, \emph{i.e.}
\begin{align}
    \frac{\vec{I}}{I_c} + \vec{i}_n &=  \overrightarrow{\sin \left(\varphi (\tau) \right)} + \overrightarrow{\frac{d \varphi (\tau)}{d \tau}},
    \label{eq:current-phase}
\end{align}{}
where $\overrightarrow{\sin \varphi}$ is a vector with components $\sin (\varphi_k)$, $\tau=\omega_c\cdot t$ is the normalized time with $\omega_c=2\pi R I_c /\Phi_0$ the characteristics frequency, and $\vec{i}_n = \vec{I}_n/I_c$ is the normalized noise current vector. 

Finally, combining Eqs. (\ref{eq:I_kirch}), (\ref{eq:phase-current2}) and (\ref{eq:current-phase}), we obtain a coupled system of first-order non-linear differential equations for $\varphi_k (\tau)$ that describes the time evolution of the array as
\begin{align}
    \overrightarrow{\frac{d \varphi}{d \tau}} = \left[\vec{i}_n - \overrightarrow{\sin (\varphi)} + \frac{\Phi_0}{2\pi I_c}  \hat{K} \hat{L}^{-1} \hat{D}\vec{\varphi} + \vec{C} \right],
    \label{eq:ODE}
\end{align}{}
where $\vec{C} = \left( \hat{K}_I^{-1}\vec{I}_b - \hat{K} \hat{L}^{-1} \vec{\Phi}_{nf} \right)/I_c$ is a vector with time-independent components. Equation (\ref{eq:ODE}) is the key equation of our paper.

\subsection{Thermal noise and numerical method}

To generate the individual thermal noise at each JJ we use the approach used by \citet{Tesche1977} and \citet{Voss1981}.
The normalized noise currents are generated at each time-step using random number generators that follow a Gaussian distribution where its mean and mean-square-deviation satisfy
\begin{align}
    \overline{i_{n,k}} &= 0, \nonumber \\
    \overline{i^2_{n,k}} &= 2\Gamma / \Delta \tau. \nonumber
\end{align}

Here $\Gamma$ is the thermal noise strength
\begin{equation}
    \Gamma= \frac{2\pi k_B T}{ \Phi_0 I_c},
    \label{eq:Gamma}
\end{equation}
where $k_B$ is the Boltzmann constant, $T$ the device operating temperature and $\Delta \tau$ the normalized time-step used when solving Eq. (\ref{eq:ODE}) numerically. 
In this work we use $T=77$ K and $\Delta \tau = 0.1$.

We have solved Eq. (\ref{eq:ODE}) using numerical integration; to do so one must choose the initial conditions of the JJ phase differences $\varphi_k(0)$. 
We found that a good choice for the initial condition of the overdamped system of Eq. (\ref{eq:ODE}) is
\begin{equation}
    \varphi_{k+1}(0) = \varphi_{k}(0) + \frac{2\pi \Phi^a_s}{\Phi_0},
    \label{eq:IC}
\end{equation}
where $\varphi_{1+N_p(n-1)}(0)=0$ for the first JJ of each row $n$.
The Euler method and Runge-Kutta 4th order method were both tested and gave convergent results. The data presented in this paper have been obtained using the Euler method since it was computationally faster. 

\subsection{Voltage of 2D arrays}
Once Eq. (\ref{eq:ODE}) is solved, we can integrate the second Josephson equation, \emph{i.e.} $V_k(t)=\frac{\Phi_0}{2\pi}\frac{\partial \varphi_k(t)}{\partial t}$, to obtain the time-averaged voltage $ \bar{V}_k $ at the $k^{\text{th}}$ JJ. Then, the normalised time-averaged voltage is $\bar{v}_k=\bar{V}_k/(I_cR)$.

The normalised time-averaged voltage $\bar{v}$ across the whole array, between top and bottom leads, is given by
\begin{equation}
    \bar{v} = \sum_{n=0}^{N_s-1} \frac{1}{N_p}\sum_{k=nN_p+1}^{(n+1)N_p}\bar{v}_k.
    \label{eq:v_arr}
\end{equation}

The time-averaged voltages across JJs in the same row are identical. Averaging over voltages in the same row as in Eq. (\ref{eq:v_arr}) improves the numerical accuracy.
In this work $10^5$ time-iterations were needed to achieve a voltage numerical error of less than $1\%$. In case of the transfer function, which is the derivative of the voltage with respect of the applied flux, smoother voltage curves are needed in order to achieve sufficient accuracy. Thus, when calculating the transfer function, $10^6$ time iterations were used.

\section{Simulation, results and discussion}\label{sec:Sim}

The time-averaged normalised voltage $\bar{v}$ of a $(N_s, N_p)$-array with uniform bias current injection depends on the parameter set $\{ N_s, N_p, I_c, \tilde{L}, T, I_b, \vec{\Phi}_a \}$ where $\tilde{L}$ represents all the partial inductances. 
The parameter set is particularly large for SQIF arrays where the SQUID loops have different sizes.
The operating temperature $T$ is taken as fixed, while $I_b$ and $\vec{\Phi}_a$ depend on external sources and can be easily adjusted. 
The bias current $I_b$ can be tuned to optimise the voltage modulation depth, $\Delta \bar{v}=\max (\bar{v}) - \min (\bar{v})$, and $\vec{\Phi}_a$ can be adjusted to find the maximum transfer function $\bar{v}_{\phi}^{\max}=\max (\partial \bar{v} /\partial \phi)$ for a given $I_b$.
$\bar{v}_{\phi}^{\max}$ can then be optimised by finding the optimal $I_b$.
If the mutual inductive coupling between SQUID loops can be neglected, the set of parameters for $\bar{v}_{\phi}^{\max}$ of a SQUID array reduces to $\{ N_s, N_p, I_c, L_s, I_b \}$ where $L_s$ is the SQUID loop self-inductance.
In this case, the alternative set  $\{ N_s, N_p, \beta_L, \Gamma, i_b \}$ can be used, where $\beta_L$ is the screening parameter, $\beta_L=2L_s I_c/\Phi_0$, and $i_b=I_b/I_c$.

In this study we consider arrays with film thickness $d=0.22$ $\upmu$m, junction width $w=2$ $\upmu$m, London penetration depth $\lambda=0.4$ $\upmu$m and bias lead lengths $100$ $\upmu$m, and the device operating temperature is fixed at $T=77$ K. Also, unless stated otherwise, we use square-SQUID loops $a_x=a_y=10$ $\upmu$m and a critical current of $I_c=20 $ $\upmu$A which is commonly found for HTS JJs \cite{Mitchell2016, Mitchell2019}. 
These values give  $\beta_L = 0.7$ and $\Gamma=0.16$ (Eq. (\ref{eq:Gamma})). 
In this paper we calculate the inductance by assuming homogeneous current density across the superconducting tracks. For the geometric partial inductances we apply the analytical expressions derived by \citet{Hoer1965}.
If more accurate inductance calculations are needed in the case of wider tracks, one can obtain the inductances using finite element methods such as 3D-MLSI \cite{Khapaev2001}, or FastHenry \cite{Kamon1994, Tausch1999} and implement them in our model.

In the $\beta_L<<1$ limit, our model shows an excellent agreement with the analytical formula given by \citet{Oppenlander2000} for 1D SQUID arrays.

\subsection{The importance of thermal noise}
\label{ssec:thermal}

YBCO step-edge JJs at $T=77$ K typically have a critical current of $I_c=20$ $\upmu$A \cite{Mitchell2016, Mitchell2019} and thus $\Gamma=0.16$. In our calculations we can turn off the effect of the thermal noise by setting $\Gamma=0$.

\begin{figure}
    \centering
    \includegraphics{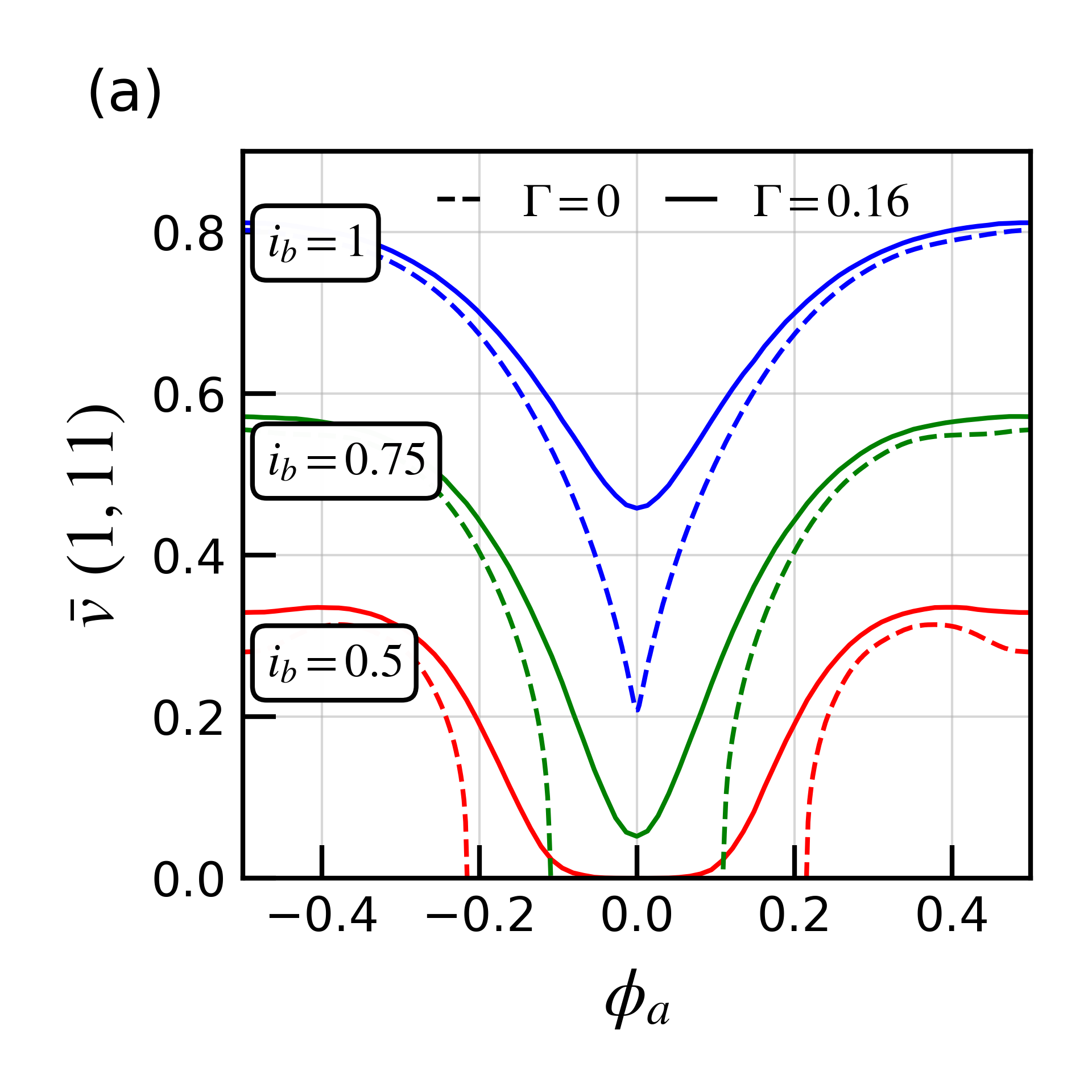}
    \includegraphics{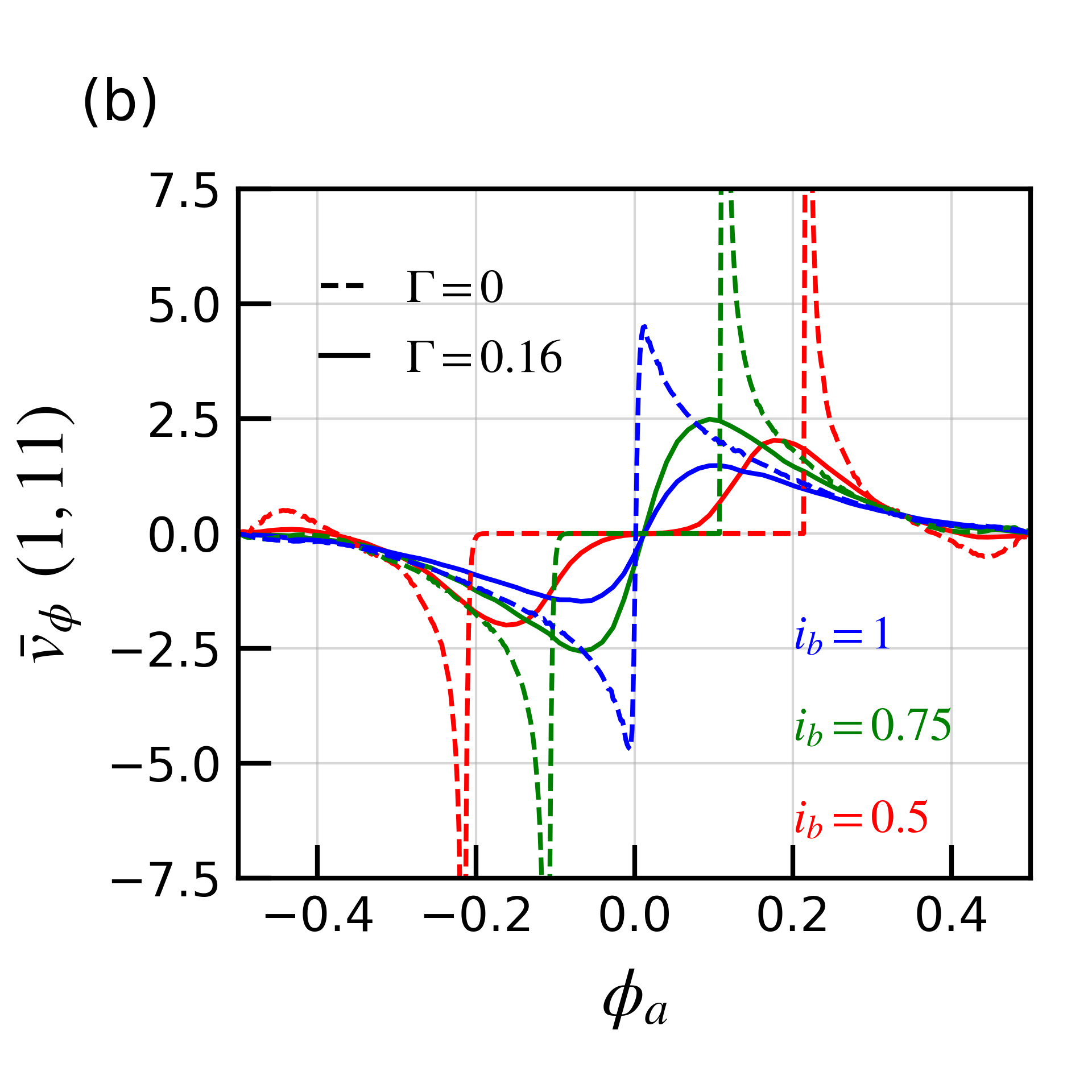}
    \caption{(a) Time averaged voltage $\bar{v}$ versus $\phi_a$ of a 1D SQUID array with $N_p=11$ at three different bias currents $i_b$ and thermal noise strength $\Gamma = 0$ (dashed lines) and at $\Gamma = 0.16$ (solid lines). (b) Corresponding transfer function $\bar{v}_{\phi}$ versus $\phi_a$.}
    \label{fig:v-vs-phi_a_T-0-77K}
\end{figure}

Figure \ref{fig:v-vs-phi_a_T-0-77K}(a) compares $\bar{v}(\phi_a)$ curves, where $\phi_a=\Phi_a/\Phi_0$, of a $(1, 11)$-SQUID array for $\Gamma=0.16$ with curves for $\Gamma=0$ at $i_b=0.5,$ 0.75 and 1. $\bar{v}(\phi_a)$ of a 1D or 2D SQUID array is  periodic in $\phi_a$ with period 1 like in the case of a symmetric dc-SQUID. 
For $i_b<1$ the dashed $\Gamma=0$ curves show zero voltage regions while with thermal noise the SQUID array is always in a non-zero voltage state for $i_b>0$. 
In Fig. \ref{fig:v-vs-phi_a_T-0-77K}(b) we show the corresponding transfer function $\bar{v}_{\phi}(\phi_a)=\partial \bar{v}(\phi_a)/\partial \phi_a$, which demonstrates that thermal noise strongly decreases the transfer function $\bar{v}_{\phi}$. Thus, it is crucial to include the effect of thermal noise when calculating the behaviour of 2D SQUID arrays at 77 K for typical $I_c$ values.

\subsection{Voltage versus magnetic flux response of $(N_s, N_p)$-SQUID arrays}
\label{ssec:2DSQUID}

\begin{figure}
    \centering
    \includegraphics[scale=1]{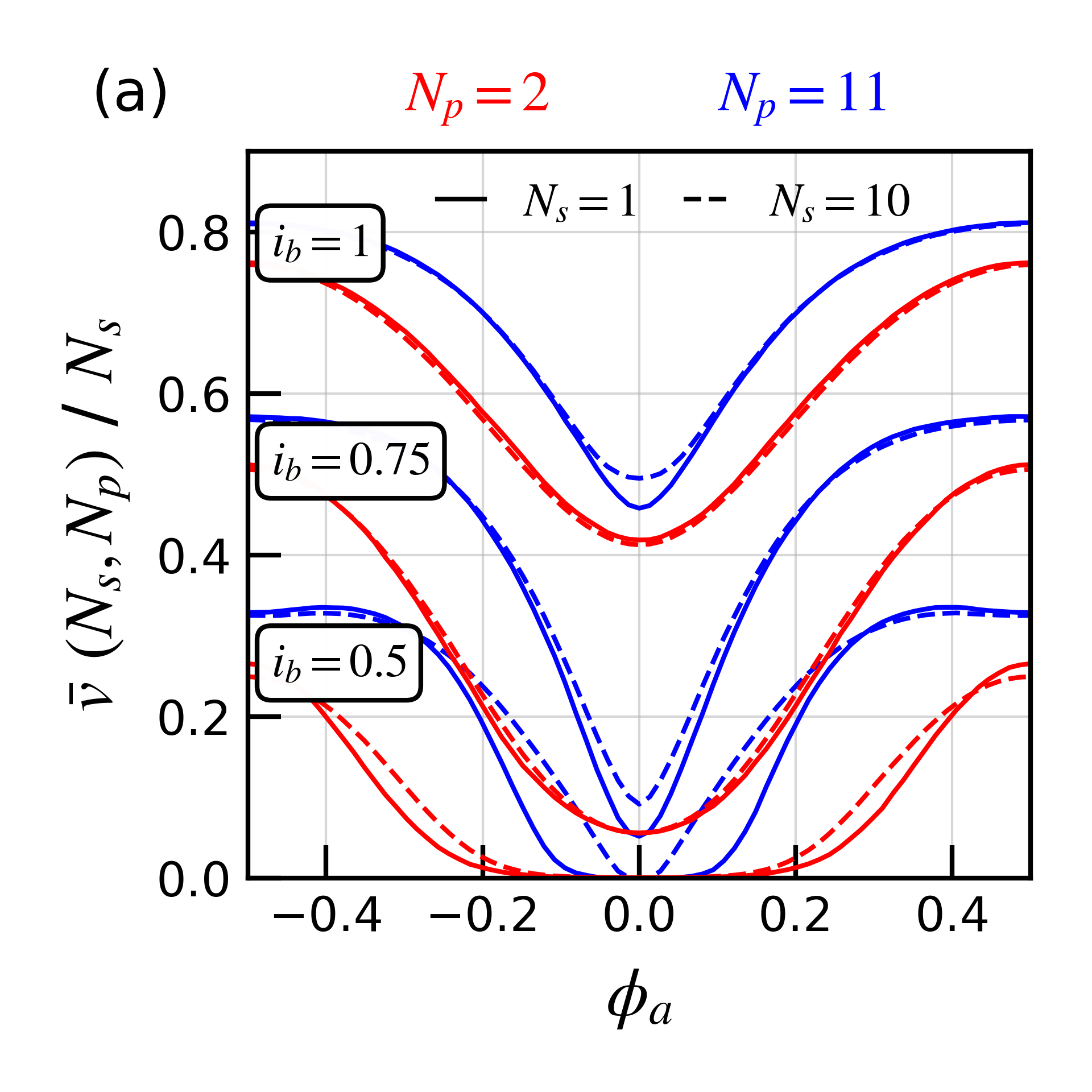} 
    \includegraphics[scale=1]{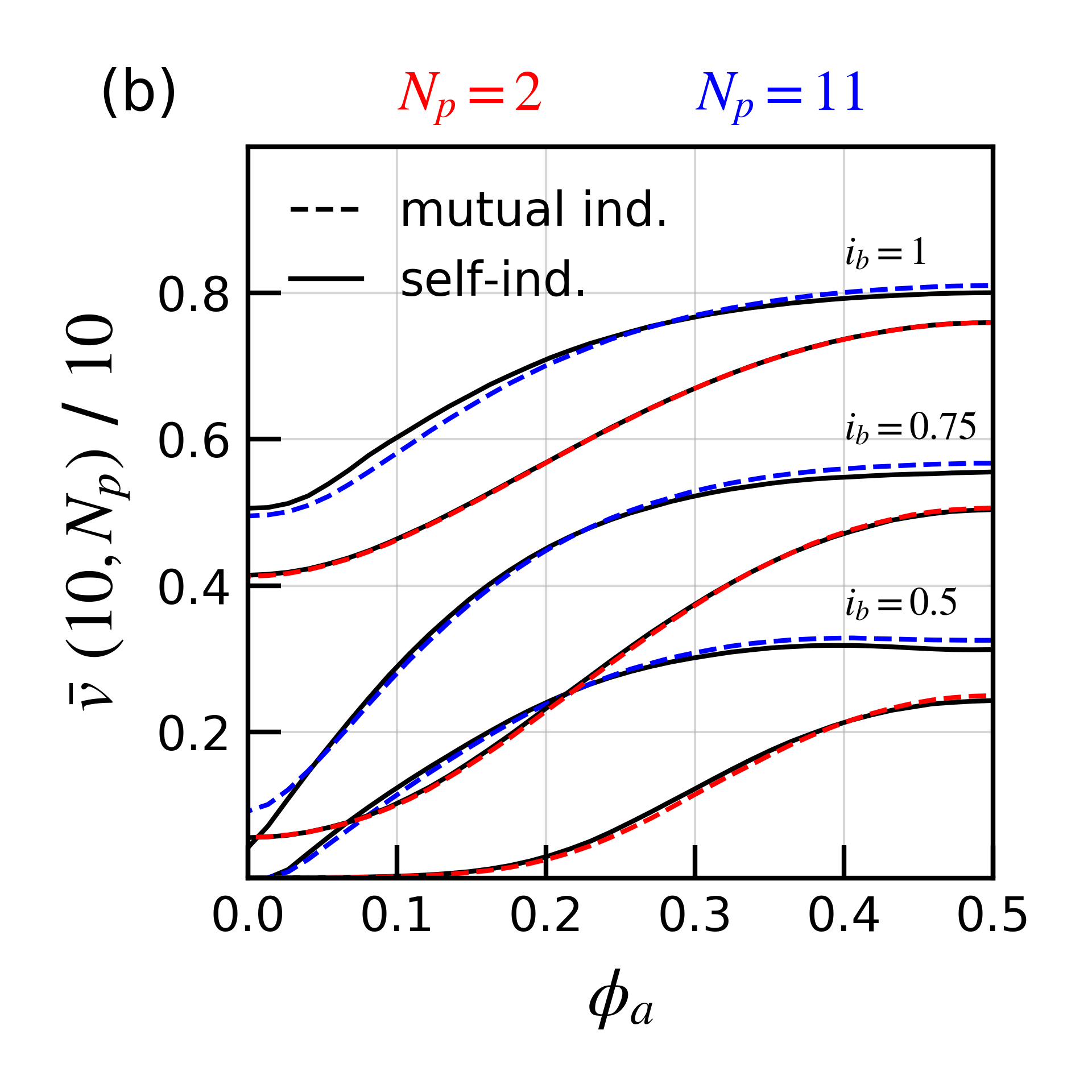} 
    \caption{(a) $\bar{v}/N_s$ versus $\phi_a$ for different bias currents $i_b$ at $T=77$ K. Four different SQUID arrays are shown. In blue SQUID arrays with $N_p=11$, in red SQUID arrays with $N_p=2$. Solid lines represent 1D arrays ($N_s=1$) and dashed lines represent 2D arrays with $N_s=10$. (b)  $\bar{v}/N_s$ versus $\phi_a$ with and without mutual inductances for $(10, N_p)$-arrays with $N_p=2$ (red) and $N_p=11$ (blue). The black solid curves are without mutual inductances, labelled ``self-ind.", while the dashed curves are with mutual inductances, labelled ``mutual ind".}
    \label{fig:2DSQUID-vs-phi_IbIc}
\end{figure}

Figure \ref{fig:2DSQUID-vs-phi_IbIc}(a) shows the $N_s$-normalised voltage $\bar{v}/N_s$ versus $\phi_a$ at different $i_b$ for four different $(N_s, N_p)$-SQUID arrays, \emph{i.e.} for $(1, 2)$, $(1, 11)$ $(10, 2)$ and $(10, 11)$. The $\bar{v}/N_s$ of narrow arrays with $N_p=2$ are displayed in red and the wider arrays with $N_p=11$ in blue.
The solid curves belong to the short arrays (1D parallel arrays) with $N_s=1$ while the dashed curves are the long arrays with $N_s=10$.
Figure \ref{fig:2DSQUID-vs-phi_IbIc}(a) reveals the validity of the scaling approximation

\begin{equation}
    \bar{v}(N_s, N_p) \approx N_s \times \bar{v}(1, N_p) .
    \label{eq:scaling_approx}
\end{equation}

When comparing the dashed curves with the solid ones in Fig. \ref{fig:2DSQUID-vs-phi_IbIc}(a), one can see that the scaling approximation Eq. (\ref{eq:scaling_approx}) holds reasonably well for certain $i_b$ and $\phi_a$, in particular for not too small $i_b$ and $\phi_a$ values.
We will discuss the validity of the scaling approximation in more detail further below.
Comparing the $N_p=2$ with the $N_p=11$ curves shows that the wider arrays produce sharper voltage dips and thus the applied flux $\phi_a^*$ that maximises the transfer function is smaller for the wider arrays.

Figure \ref{fig:2DSQUID-vs-phi_IbIc}(b) illustrates for $N_s=10$ the effect of the mutual inductances on $\bar{v}(\phi_a)$ for $N_p=2$ and 11 at three different $i_b$. Here $\beta_L=0.7$ and $\Gamma=0.16$. The dashed lines are with mutual inductances while the black solid lines without them. Only minor differences are noticeable for the overall voltage response.
Interestingly, \citet{Dalichaouch2014} claimed that mutual inductances are important to obtain the correct transfer function and voltage response, but their calculations have been done without thermal noise at $\Gamma=0$.
The computational time needed for our simulations did not increase significantly when mutual inductances were included, and therefore the simulations presented in this paper are with mutual inductances.

\subsection{Maximum transfer function dependence on the bias current}
\label{ssec:v_phi-ib}

\begin{figure}[h!]
    \centering
    \includegraphics{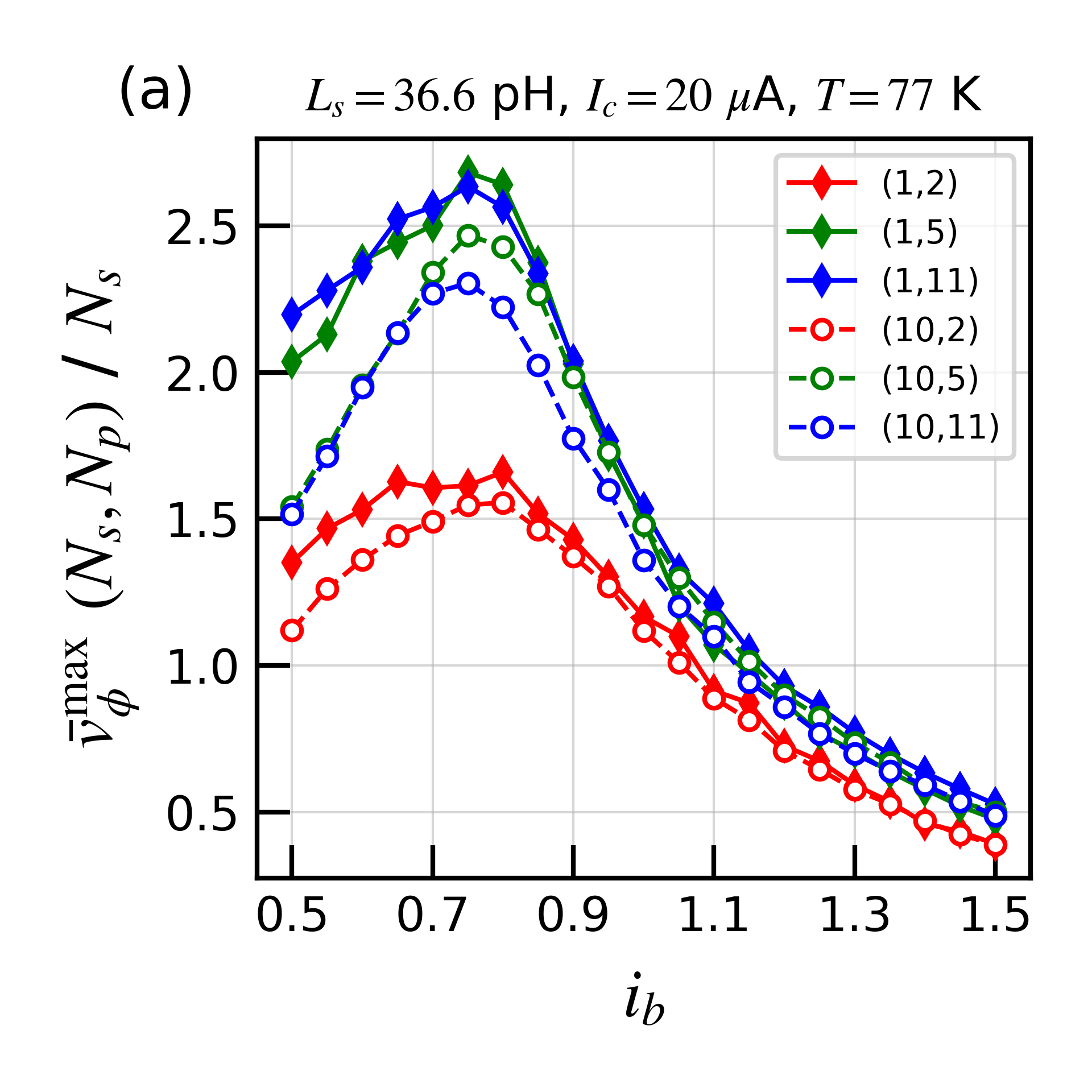}
    \includegraphics{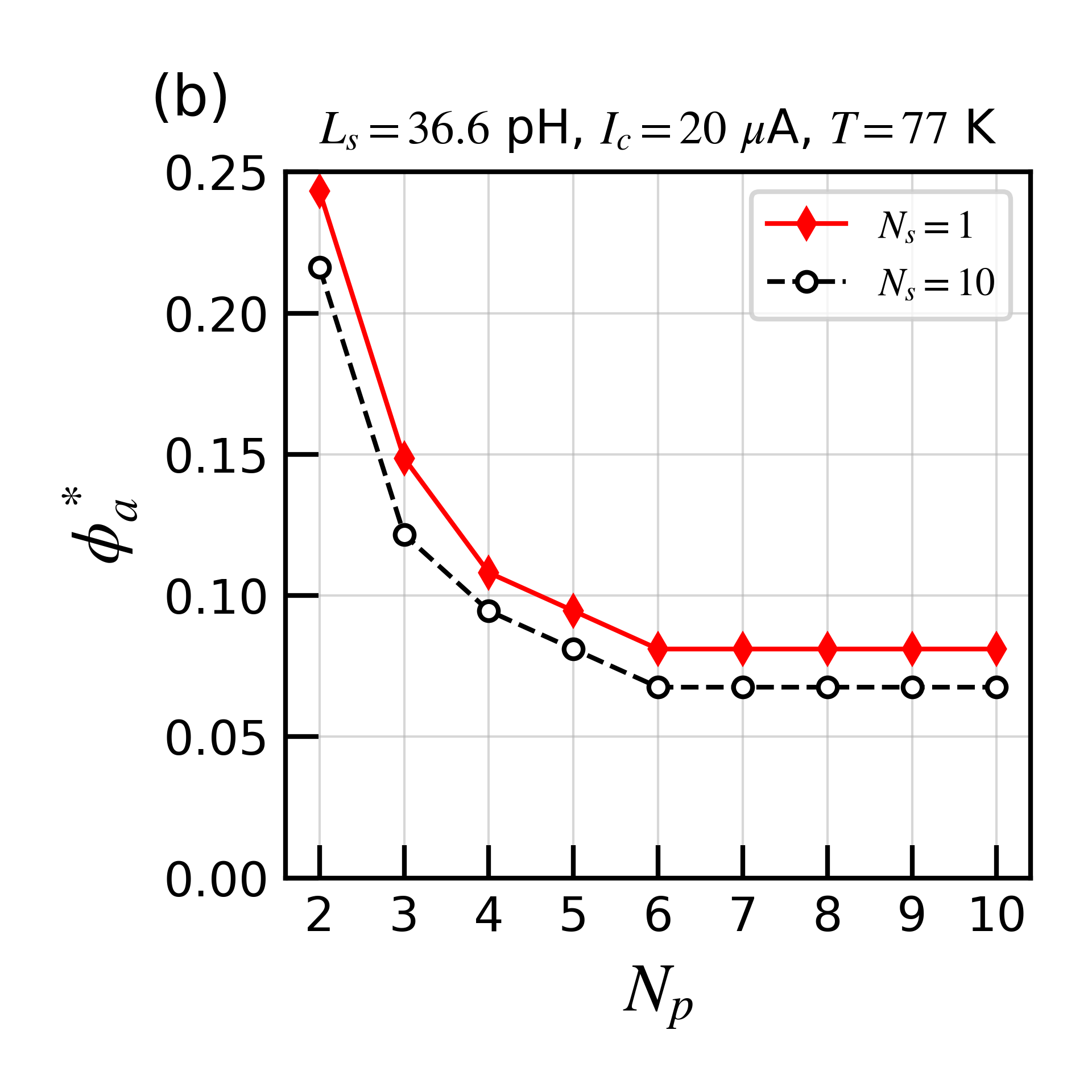}
    \caption{(a) Normalised maximum transfer function $\bar{v}_{\phi}^{\max} / N_s$ versus bias current $i_b$ at $T=77$ K for six different SQUID arrays, \emph{i.e.} three 1D arrays (solid lines with diamond symbols) and three 2D arrays with $N_s=10$ (dashed lines with circles). Colors indicate the number of JJs in parallel: $N_p=2$ (red), $N_p=5$ (green) and $N_p=11$ (blue). (b) Applied magnetic flux $\phi_a^*$ that maximises $\bar{v}_{\phi}$ versus $N_p$ for $(1, N_p)$-SQUID arrays (solid lines with diamonds) and $(10, N_p)$-SQUID arrays (dashed lines with circles) at 77 K for the optimal bias current $i_b^{opt}=0.75$.}
    \label{fig:dvdphi-vs-ib}
\end{figure}

In Fig. \ref{fig:dvdphi-vs-ib}(a) we show the normalised maximum transfer function $\bar{v}_{\phi}^{\max}/N_s=\max(\partial \bar{v} / \partial \phi_a)/N_s$ versus the bias current, $i_b=I_b/I_c$, for six different $(N_s, N_p)$-SQUID arrays operating at $T=77$ K. 
The solid lines with diamond symbols correspond to 1D SQUID arrays and the dashed lines with circles describe 2D SQUID arrays with $N_s=10$. The colours describe the number of junctions in parallel of each array, \emph{i.e.} red for $N_p=2$, green for $N_p=5$ and blue for $N_p=11$. 
From Fig. \ref{fig:dvdphi-vs-ib}(a) we can see that an optimal bias current $i_b^{opt}$ exists for each array with $i_b^{opt}\approx 0.75$.
The figure also reveals that the scaling approximation (Eq. \ref{eq:scaling_approx}) is valid for the $\bar{v}_{\phi}^{\max}$ within about $ 20 \%$ at bias currents close to the $i_b^{opt}$. For larger bias currents the maximum transfer functions scale very well with $N_s$ for all $N_p$ studied.
Figure \ref{fig:dvdphi-vs-ib}(a) further shows a significant increase of $\bar{v}_{\phi}^{\max}/N_s$ from $N_p=2$ to 5. This is due to the sharpening of the dip of $\bar{v}(\phi_a)$ with $N_p$, which could also be seen in Fig. \ref{fig:2DSQUID-vs-phi_IbIc}(a) comparing the $N_p=2$ arrays with the $N_p=11$ ones. This effect has previously been reported by \citet{Oppenlander2000} for uniformly biased SQIF arrays at $T=0$ K.
While in Fig. \ref{fig:dvdphi-vs-ib}(a) there is a large difference between $N_p=2$ and 5, this is not the case for $N_p=5$ and 11. The reason for this will be discussed below.

The applied flux $\phi_a^*$ that maximises the transfer function $\partial \bar{v} / \partial \phi_a$ at $i_b^{opt}=0.75$ is shown in Fig. \ref{fig:dvdphi-vs-ib}(b) as a function of $N_p$ for $N_s=1$ and 10. While $\phi_a^*\approx 0.25$ for a dc-SQUID and dc-SQUID in series, $\phi_a^*$ drops to $\phi_a^*\approx 0.07$ for wider arrays with $N_p \geq 5$. There is not much difference in $\phi_a^*$ for $N_s=1$ and $N_s=10$.

\subsection{Maximum transfer function dependence on critical current and SQUID self-inductance}
\label{ssec:v_phi_L_Ic}

Previously we have seen that the contribution of the mutual inductance for these kind of arrays is negligible, and therefore either $L_s$ and $I_c$ or $\beta_L$ and $\Gamma$ can be used as array parameters.

\begin{figure}[h!]
    \centering
    \includegraphics{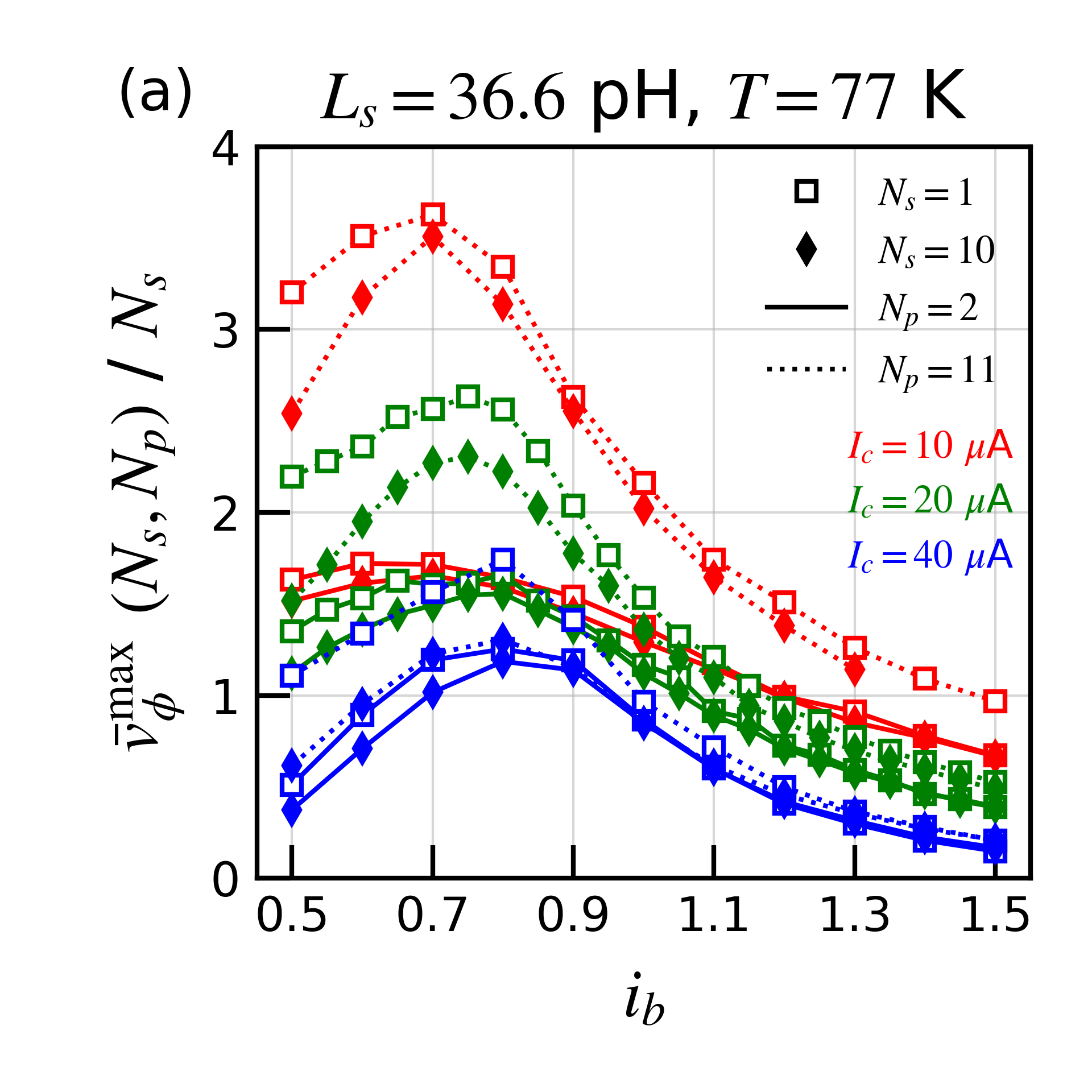}
    \includegraphics{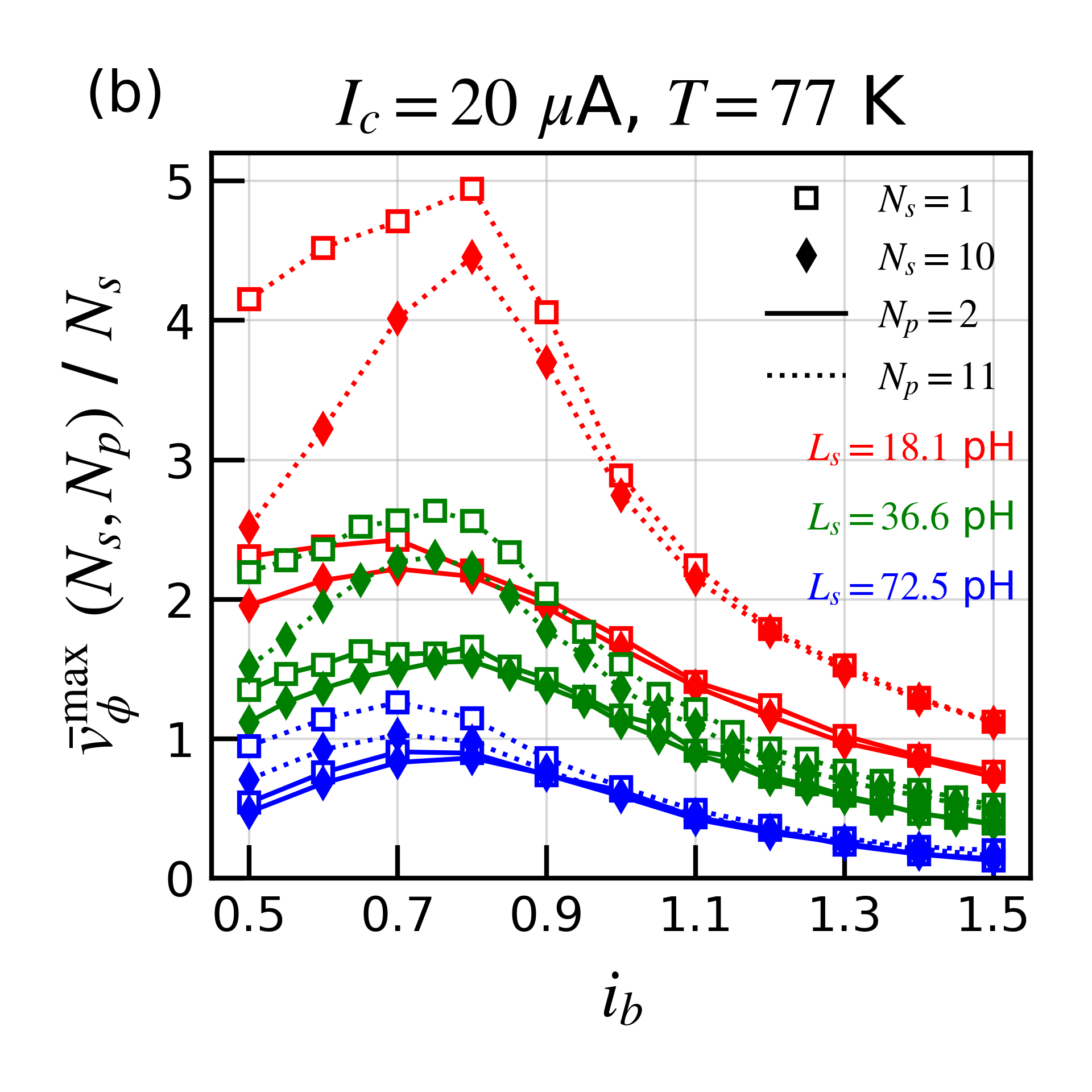}
    \caption{Normalised maximum transfer function $\bar{v}_{\phi}^{\max}/N_s$ versus bias current $i_b$ at $T=77$ K of 1D SQUID arrays (squares) and 2D SQUID arrays (diamonds) with $N_p=2$ (solid lines) and $N_p=11$ (dotted lines). (a) Arrays with different critical currents $I_c$. The SQUID inductance is fixed at $L_s=36.6$ pH. In red  $I_c=10$ $\upmu$A ($\beta_L=0.35$, $\Gamma=0.32$), in green $I_c=20$ $\upmu$A ($\beta_L= 0.7$, $\Gamma=0.16$), and in blue $I_c=40$ $\upmu$A ($\beta_L=1.4$, $\Gamma=0.08$). (b) Arrays with different SQUID self-inductances $L_s$. The JJ critical current is fixed at $I_c=20$ $\upmu$A ($\Gamma=0.16$). In red $L_s=18.1$ pH ($\beta_L=0.35$), in green  $ L_s=36.6$ pH ($\beta_L = 0.7$); and in blue $L_s=72.5$ pH ($\beta_L=1.4$).}
    \label{fig:dvdphi-vs-ib_var-IcL}
\end{figure}

Using $L_s=36.6$ pH, which corresponds to a SQUID loop size of $a_x=a_y=10$ $\upmu$m, in Fig. \ref{fig:dvdphi-vs-ib_var-IcL}(a) we show the normalised maximum transfer function $\bar{v}_{\phi}^{\max}/N_s$ versus the bias current $i_b$ for arrays with different $I_c$. Colours depict different critical currents, \emph{i.e.} in red for $I_c=10$ $\upmu$A ($\beta_L=0.35$, $\Gamma=0.32$), in green for $I_c=20$ $\upmu$A ($\beta_L=0.7$ and $\Gamma=0.16$), and in blue for $I_c=40$ $\upmu$A ($\beta_L=1.4$ and $\Gamma=0.08$). The line-style defines the array size, \emph{i.e.} square symbols for $N_s=1$ and diamond symbols for $N_s=10$, while the solid lines represent arrays with $N_p=2$ and dotted lines for $N_p=11$.
Fig. \ref{fig:dvdphi-vs-ib_var-IcL}(a) shows that decreasing $I_c$ increases  $\bar{v}_{\phi}^{\max}/N_s$ for all $(N_s, N_p)$ combinations while the optimal bias current $i_b^{opt}$ decreases from $i_b^{opt}=0.8$ to 0.7.
The scaling approximation Eq. (\ref{eq:scaling_approx}) holds very well for arrays with $N_p=2$, independently of $I_c$, while for wider arrays, $N_p=11$, it seems to worsen with increasing $I_c$ values.

Figure \ref{fig:dvdphi-vs-ib_var-IcL}(b) illustrates the dependence of $\bar{v}_{\phi}^{\max}(i_b)/N_s$ on the SQUID loop self-inductance $L_s$ for a fixed $I_c=20$ $\upmu$A. The line-style in Fig. \ref{fig:dvdphi-vs-ib_var-IcL}(b) indicates the $(N_s, N_p)$ numbers, while the colour differentiates between the SQUID self-inductance $L_s$ values.
The red curves correspond to $L_s=18.1$ pH ($\beta_L=0.35$) for which $a_x=a_y= 5.6$ $\upmu$m, the green curves to $L_s=36.6$ pH ($\beta_L=0.7$) for which $a_x=a_y= 10$ $\upmu$m and the blue ones to $L_s=72.4$ pH ($\beta_L=1.4$) for which $a_x=a_y= 17.7$ $\upmu$m.
In contrast to the previous Fig. \ref{fig:dvdphi-vs-ib_var-IcL}(a) where both $\beta_L$ and $\Gamma$ changed due to the change in $I_c$, here only $\beta_L$ changes while $\Gamma$ stays fixed at $\Gamma=0.16$.

As it can be seen from Fig. \ref{fig:dvdphi-vs-ib_var-IcL}(b), $\bar{v}_{\phi}^{\max}(i_b)/N_s$ increases with decreasing the SQUID loop self-inductance $L_s$.
As in the previous Fig. \ref{fig:dvdphi-vs-ib_var-IcL}(a), the scaling approximation, Eq. (\ref{eq:scaling_approx}), again holds fairly well, especially for the $N_p=2$ case (dc-SQUIDs in series).
Figures \ref{fig:dvdphi-vs-ib_var-IcL}(a) and \ref{fig:dvdphi-vs-ib_var-IcL}(b) reveal that the scaling approximation is not sensitive to the above choices of $I_c$ and $L_s$.

\subsection{Maximum transfer function dependence on $N_s$, $N_p$ and the coupling radius}
\label{ssec:v_phi-3D}

\begin{figure}[h!]
    \centering
    \includegraphics{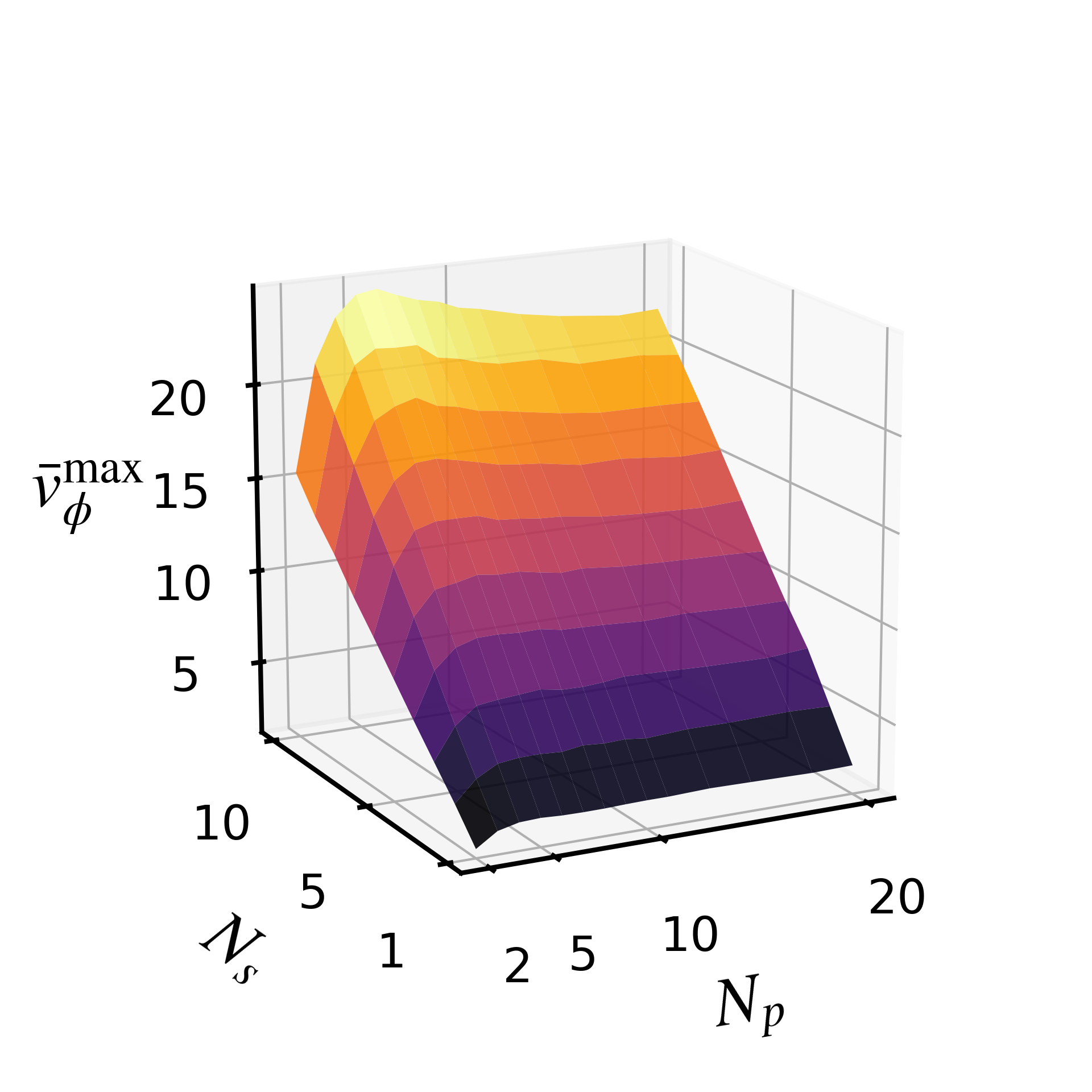}
    \caption{Maximum transfer function $\bar{v}_{\phi}^{\max}$ versus $N_s$ and $N_p$ at $T=77$ K for a bias current $i_b=0.75$. The SQUID arrays have square SQUID loops, \emph{i.e.} $a_x=a_y=10$ $\upmu$m ($L_s=36.6$ pH), $I_c=20$ $\upmu$A ($\beta_L=0.7$ and $\Gamma=0.16$), and $N_p^*$ is about 5.}
    \label{fig:3d-map}
\end{figure}

To understand better the dependence of the maximum transfer function $\bar{v}_{\phi}^{\max}$ on the array dimensions, $N_s$ and $N_p$, at $T=77$ K we show in Fig. \ref{fig:3d-map} a three-dimensional plot using a bias current of $i_b=0.75$, which is close to the optimal bias current for most arrays.
Figure \ref{fig:3d-map} clearly shows the linear dependence on $N_s$. In contrast, $\bar{v}_{\phi}^{\max}$ initially increases with $N_p$ up to $N_p^* \approx 5$, followed by a small decrease and a plateau for larger $N_p$. $N_p^*$ does not depend on $N_s$. 
This plateauing behaviour was previously reported for 1D parallel array calculations at $T=0$ K by Kornev \emph{et al.}\cite{Kornev2009b, Kornev2011} and also by \citet{Mitchell2019}. Kornev \emph{et al.}\cite{Kornev2009b, Kornev2011} explained this behaviour by the concept of a ``coupling radius" (or ``interaction radius") $N_p^*$, which arises because the array acts as an R-L network, which, depending on its operating frequency, only allows $N_p^*$ parallel JJs to couple or interact.
According to Kornev \emph{et al.}\cite{Kornev2011}, $N_p^*$ depends on the normalised coupling impedance $\omega l$ of the array which can be expressed as $\omega l = \pi \beta_L \bar{v}(i_b^{opt}, \phi_a^*)$. 

Importantly, Fig. \ref{fig:3d-map} reveals that $\bar{v}_{\phi}^{\max}(N_s, N_p)$ can be approximated by
\begin{equation}
    \bar{v}_{\phi}^{\max} (N_s, N_p) \approx \frac{N_s \tilde{N}_p}{2} \times \bar{v}_{\phi}^{\max} (1, 2),
    \label{eq:max-v-phi_approx}
\end{equation}
with $\tilde{N}_p=N_p$ if $N_p<N_p^*$ and $\tilde{N}_p=N_p^*$ if $N_p \geq N_p^*$, where $\bar{v}_{\phi}^{\max}(1,2)$ is the maximum transfer function of the dc-SQUID. $N_p^*$ increases with decreasing the self-inductance $L_s$ and $N_p^*\rightarrow \infty$ if $L_s \rightarrow 0$.

\subsection{Voltage modulation depth dependence on bias current}
\label{ssec:dv}

\begin{figure}[!h]
    \centering
    \includegraphics{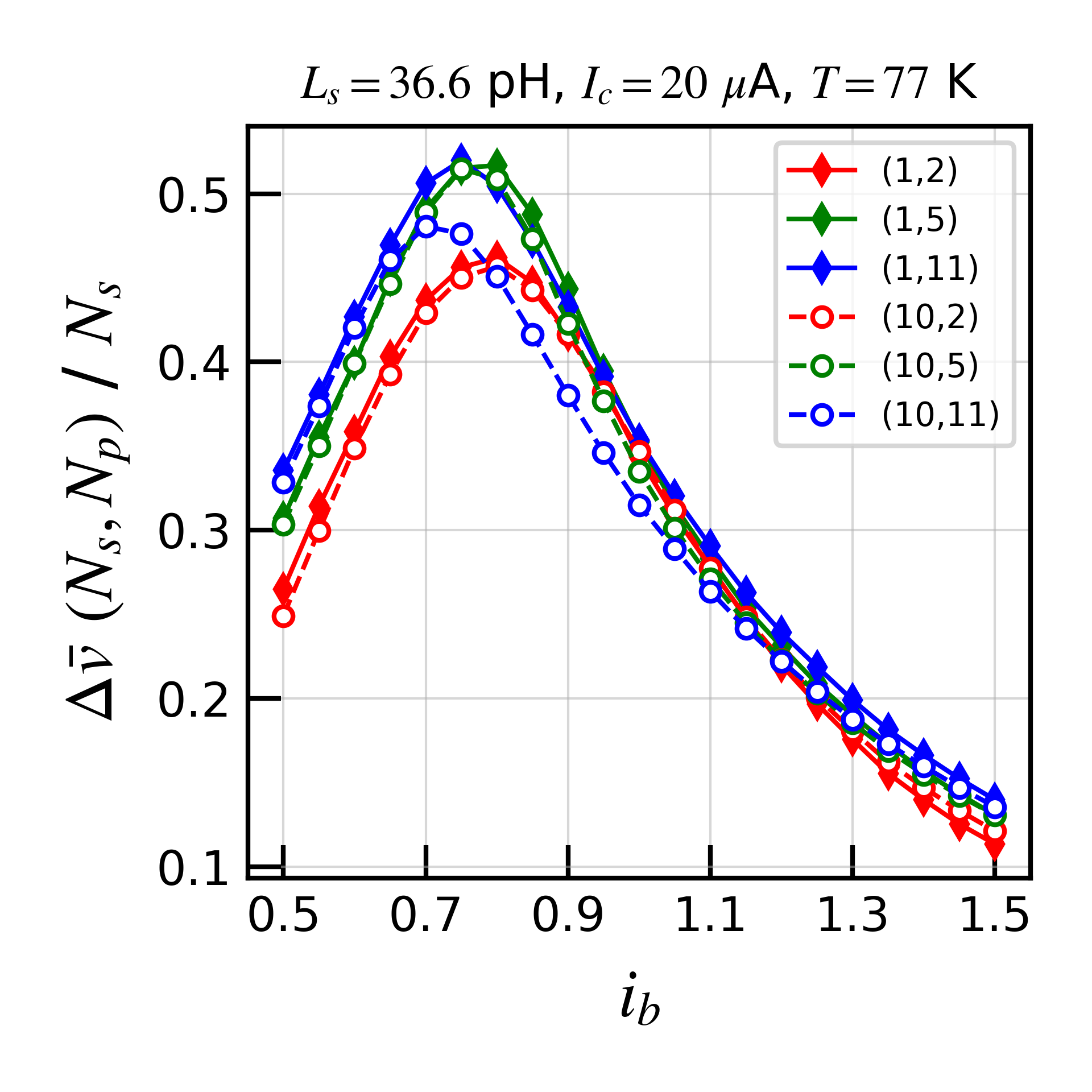}
    \caption{Normalised voltage modulation depth $\Delta \bar{v} / N_s$ versus bias current $i_b$ at $T=77$ K for three 1D arrays (solid lines with diamonds) and three 2D arrays with $N_s=10$ (dashed lines with circles). Colors indicate the number of JJs in parallel: $N_p=2$ (red), $N_p=5$ (green) and $N_p=11$ (blue).}
    \label{fig:dv-vs-ib}
\end{figure}

In Fig. \ref{fig:dv-vs-ib} we study the same arrays as in Fig. \ref{fig:dvdphi-vs-ib}(a), but in this case instead of the $\bar{v}_{\phi}^{\max}$ we analyse the normalised voltage modulation depth $\Delta \bar{v}/N_s= (\max(\bar{v}) - \min(\bar{v})/N_s$ versus the bias current $i_b=I_b/I_c$ operating at $T=77$ K. Solid lines with diamond symbols correspond to 1D SQUID arrays and dashed lines with circles describe 2D SQUID arrays with $N_s=10$. The colours indicate the number of JJ in parallel, \emph{i.e.} red for $N_p=2$, green for $N_p=5$ and blue for $N_p=11$. 
From Fig. \ref{fig:dv-vs-ib} we can see that an optimal bias current $i_b^{opt}$ exists for each array with $i_b^{opt} \approx 0.7-0.8$.
One can also see that the scaling approximation (Eq. \ref{eq:scaling_approx}) holds very well for $N_p=2$ and 5. For $N_p=11$ the scaling approximation holds within $10\%$ in the region close to $i_b^{opt}$. The optimal bias current for $N_p=11$ is slightly different between the 1D and 2D arrays, being $i_b^{opt} \approx 0.7$ for the 2D array and $i_b^{opt} \approx 0.75$ for the 1D one. 

Comparing Figs. \ref{fig:dvdphi-vs-ib}(a) with Fig. \ref{fig:dv-vs-ib} reveals that $i_b^{opt}$ for $\Delta \bar{v}/N_s$ and for $\bar{v}_{\phi}^{\max}/N_s$ are very similar, and therefore experimentally one could use the maximum voltage modulation depth to obtain the optimal bias current of the maximum transfer function. 

\subsection{Voltage versus magnetic flux response of $(N_s, N_p)$-SQIF arrays}
\label{ssec:SQIF}

For some applications like absolute field magnetometers \cite{Caputo2005} the periodicity of the $\bar{v}(\phi_a)$ response of dc-SQUIDs and SQUID arrays is not a desirable feature since it does not offer a unique response. 
For these kind of applications SQIF arrays are preferred because the periodicity with the magnetic flux is broken by introducing a spread in the SQUID loop areas of the array. 
The array structures considered in our model follow grid-like patterns. This kind of structure implies that SQUIDs in the same row must have same height and SQUIDs in the same column must have same width (see Fig. \ref{fig:diag}). 
With these two restrictions in mind, we can create a SQIF response by changing the width and/or height of the SQUID loops.

\begin{figure}[h!]
    \centering
    \includegraphics[scale=1]{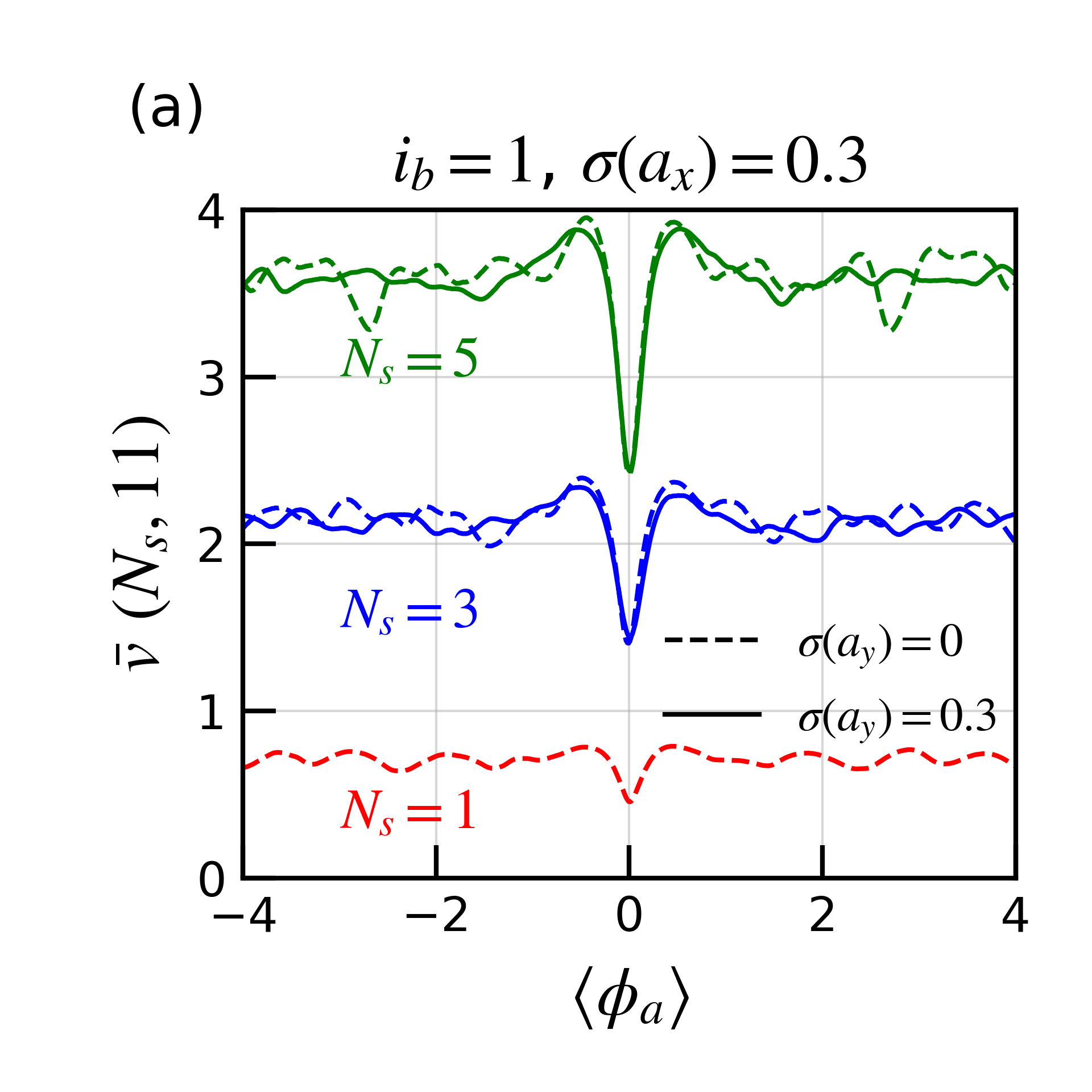}
     \includegraphics[scale=1]{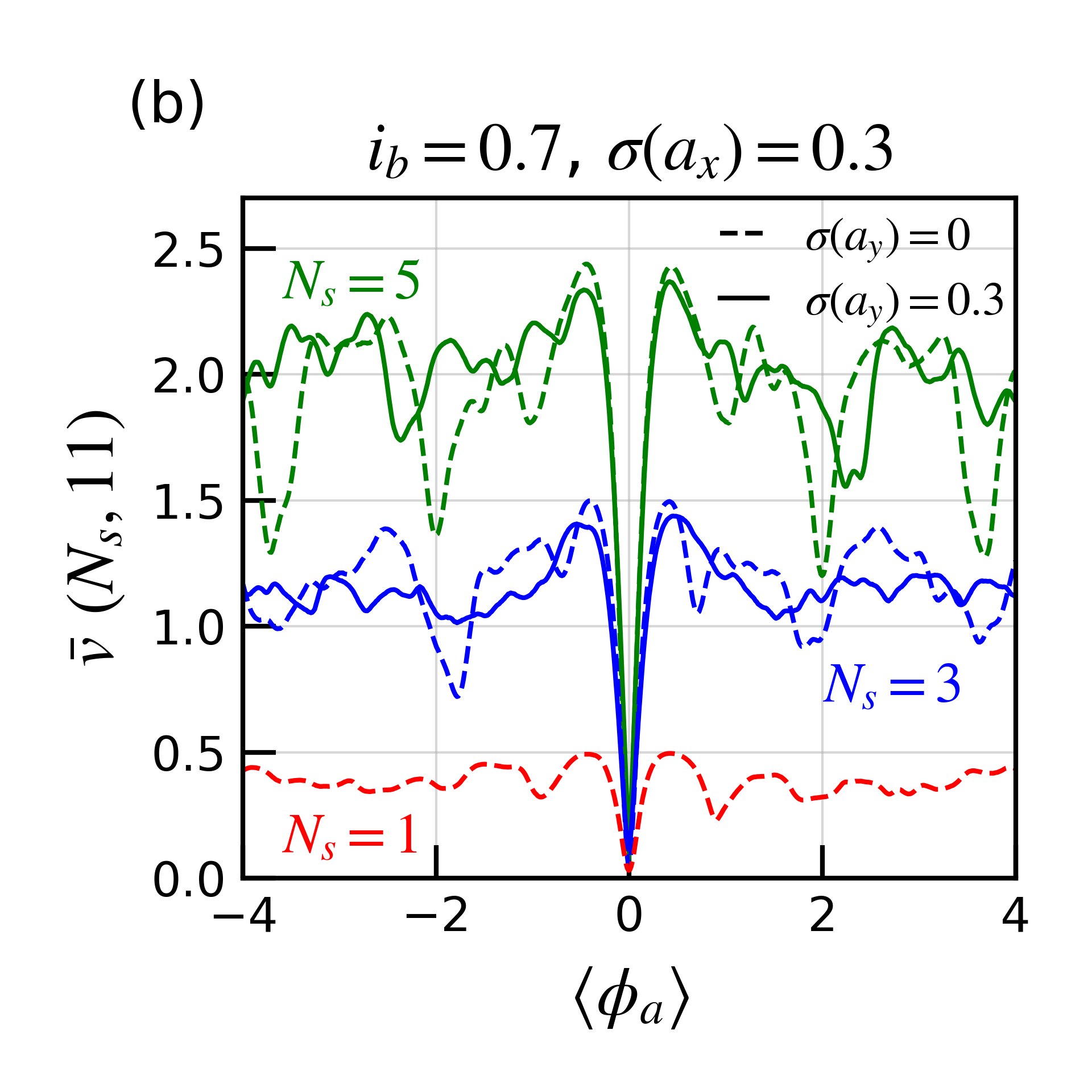}
    \caption{Time-averaged voltage $\bar{v}$ dependence on the averaged normalized applied magnetic flux $\langle \phi_a \rangle$ of SQIF arrays with $N_p=11$ for different $N_s$. In red the 1D SQIF, in blue the 2D SQIF with $N_s=3$ and in green the 2D SQIF with $N_s=5$. Dashed lines represent SQIF arrays with only an $a_x$ spread, $\sigma (a_x) =0.3$, while solid lines show 2D SQIFs with spreads in $a_x$ and $a_y$. The SQUID mean sizes are $\langle a_x \rangle = \langle a_y \rangle =10$ $\upmu$m, $T=77$ K and the bias current is $i_b = 1$ (a) and $i_b=0.7$ (b).}
    \label{fig:2DSQIF}
\end{figure}

In Fig. \ref{fig:2DSQIF}(a) we show the voltage $\bar{v}$ versus the averaged magnetic flux $\langle \phi_a \rangle=\sum_{s=1}^{N_{SQ}}\phi^a_s/N_{SQ}$ of five different SQIF arrays at $T=77$ K and $i_b=1$.
All these arrays have eleven junctions in parallel ($N_p=11$), and the colours of the curves indicate the number of SQUIDs in series, with red $N_s=1$, blue $N_s=3$ and green $N_s=5$. 
The SQIFs represented with dashed lines are created by only varying the width $a_x$ of the SQUIDs using random normal distributed values with a standard deviation of $\sim 30\%$. 
The 2D SQIFs ($N_s=3$ and 5) represented with solid lines are obtained by adding a $\sim 30 \%$ spread in the height $a_y$ of the SQUIDs. 
Figure \ref{fig:2DSQIF}(a) shows that the SQIFs with spreads in both directions have less prominent secondary peaks than the SQIFs with spread only in $a_x$. The modulation depth of the main dip gets slightly reduced for spreads in both $a_x$ and $a_y$. 

Figure \ref{fig:2DSQIF}(b) shows $\bar{v}$ versus $\langle \phi_a \rangle$ for $i_b=0.7$, close to the optimal $i_b^{opt}$. We can see that the voltage modulation depths of the main dip of the SQIF arrays in Fig. \ref{fig:2DSQIF}(b) are larger than those of Fig. \ref{fig:2DSQIF}(a), were the bias current is $i_b=1$.

In Fig. \ref{fig:2DSQIF}(b) we can see more clearly the effects of considering loop area spreads in both $a_x$ and $a_y$. By adding the spread in the loop width of each column and in the loop height of each row we achieve a larger spread in the SQUID loop areas which creates stronger destructive interference. 

Instead of generating SQIF arrays using randomly generated spreads with a given mean value, one could generate the SQUID loops in a systematic manner. 
One choice, that has been used before for 1D SQIF arrays \cite{Oppenlander2000}, is using Gaussian arrays to determine the height of each row and width of each column. A systematic method would further reduce secondary peaks since it would ensure distinct SQUID loop areas. It also could help to achieve greater linearity of $\bar{v}(\phi_a)$ around $\phi_a^*$ \cite{Kornev2009a}.

\section{Summary}
\label{sec:Con}

In this paper we presented a theoretical model which describes the behaviour of 1D and 2D SQUID and SQIF arrays for uniform bias current injection at $T = 77$ K.
Besides the parameters that characterise a single dc-SQUID here the number $N_s$ of SQUID rows and the number $N_p$ of JJs in parallel as well as the mutual inductive coupling between SQUID loops become additional parameters. The largest arrays size that we studied was $(N_s,N_p) = (10,11)$.

Our results showed that the inclusion of Johnson thermal noise is paramount for correctly predicting the voltage response and the maximum transfer function. By turning the mutual inductances on and off we elucidated the role of the inductive coupling between SQUID loops, revealing that the contribution from mutual inductances is negligibly small for these array sizes.

Furthermore, our simulations established the validity of an approximate scaling behaviour for the voltage in the form $\bar{v}(N_s,N_p) \approx N_s \times \bar{v}(1,N_p)$ for certain bias currents and applied fluxes. Such an approximate scaling was also found for the maximum transfer function and is most accurate for narrow arrays and large bias currents. The applied magnetic flux that maximises the transfer function was found to decrease with increasing $N_p$.

We demonstrated that the maximum transfer function can be be optimised with a bias current $i_b^{opt} = 0.7 - 0.8$. The optimal bias current was shown to depend only weakly on our choice of the SQUID self-inductance and JJ critical current. 

Most importantly, our simulations revealed that the maximum transfer function of a 2D SQUID array is proportional to the maximum transfer function of the corresponding dc-SQUID and scales with $N_s N_p / 2$ or $N_s N_p^* / 2$ where $N_p^*$ is the so called coupling radius. 

Our calculations also showed that the bias current which optimises the maximum transfer function also optimises the voltage modulation depth of 2D-SQUID arrays.

In addition, we studied 2D SQIF arrays where we compared arrays which only had a spread in the SQUID loop height with arrays with spreads in both SQUID loop height and width.
We showed that increasing the spread in both directions further reduces secondary oscillations in the voltage response.

Our work offers researchers a theoretical model that can accurately simulate 1D and 2D SQUID and SQIF arrays made from HTS materials as it fully includes thermal noise. The model could be used in the future to further deepen our understanding of the complicated parameter dependence of 2D SQUID and SQIF arrays.


\newpage
\bibliography{model_2D-SQIF.bib}

\begin{thebibliography}{35}%
\makeatletter
\providecommand \@ifxundefined [1]{%
 \@ifx{#1\undefined}
}%
\providecommand \@ifnum [1]{%
 \ifnum #1\expandafter \@firstoftwo
 \else \expandafter \@secondoftwo
 \fi
}%
\providecommand \@ifx [1]{%
 \ifx #1\expandafter \@firstoftwo
 \else \expandafter \@secondoftwo
 \fi
}%
\providecommand \natexlab [1]{#1}%
\providecommand \enquote  [1]{``#1''}%
\providecommand \bibnamefont  [1]{#1}%
\providecommand \bibfnamefont [1]{#1}%
\providecommand \citenamefont [1]{#1}%
\providecommand \href@noop [0]{\@secondoftwo}%
\providecommand \href [0]{\begingroup \@sanitize@url \@href}%
\providecommand \@href[1]{\@@startlink{#1}\@@href}%
\providecommand \@@href[1]{\endgroup#1\@@endlink}%
\providecommand \@sanitize@url [0]{\catcode `\\12\catcode `\$12\catcode
  `\&12\catcode `\#12\catcode `\^12\catcode `\_12\catcode `\%12\relax}%
\providecommand \@@startlink[1]{}%
\providecommand \@@endlink[0]{}%
\providecommand \url  [0]{\begingroup\@sanitize@url \@url }%
\providecommand \@url [1]{\endgroup\@href {#1}{\urlprefix }}%
\providecommand \urlprefix  [0]{URL }%
\providecommand \Eprint [0]{\href }%
\providecommand \doibase [0]{http://dx.doi.org/}%
\providecommand \selectlanguage [0]{\@gobble}%
\providecommand \bibinfo  [0]{\@secondoftwo}%
\providecommand \bibfield  [0]{\@secondoftwo}%
\providecommand \translation [1]{[#1]}%
\providecommand \BibitemOpen [0]{}%
\providecommand \bibitemStop [0]{}%
\providecommand \bibitemNoStop [0]{.\EOS\space}%
\providecommand \EOS [0]{\spacefactor3000\relax}%
\providecommand \BibitemShut  [1]{\csname bibitem#1\endcsname}%
\let\auto@bib@innerbib\@empty
\bibitem [{\citenamefont {Clarke}\ and\ \citenamefont
  {Braginski}(2004)}]{Clarke2004}%
  \BibitemOpen
  \bibfield  {author} {\bibinfo {author} {\bibfnamefont {J.}~\bibnamefont
  {Clarke}}\ and\ \bibinfo {author} {\bibfnamefont {A.~I.}\ \bibnamefont
  {Braginski}},\ }\href {\doibase 10.1002/3527603646} {\emph {\bibinfo {title}
  {Materials and Manufacturing Processes}}},\ edited by\ \bibinfo {editor}
  {\bibfnamefont {J.}~\bibnamefont {Clarke}}\ and\ \bibinfo {editor}
  {\bibfnamefont {A.~I.}\ \bibnamefont {Braginski}},\ Vol.~\bibinfo {volume}
  {1}\ (\bibinfo  {publisher} {Wiley},\ \bibinfo {year} {2004})\BibitemShut
  {NoStop}%
\bibitem [{\citenamefont {Clem}\ and\ \citenamefont {Brandt}(2005)}]{Clem2005}%
  \BibitemOpen
  \bibfield  {author} {\bibinfo {author} {\bibfnamefont {J.~R.}\ \bibnamefont
  {Clem}}\ and\ \bibinfo {author} {\bibfnamefont {E.~H.}\ \bibnamefont
  {Brandt}},\ }\href {\doibase 10.1103/PhysRevB.72.174511} {\bibfield
  {journal} {\bibinfo  {journal} {Physical Review B}\ }\textbf {\bibinfo
  {volume} {72}},\ \bibinfo {pages} {174511} (\bibinfo {year}
  {2005})}\BibitemShut {NoStop}%
\bibitem [{\citenamefont {Foley}\ \emph {et~al.}(1999)\citenamefont {Foley},
  \citenamefont {Leslie}, \citenamefont {Binks}, \citenamefont {Lewis},
  \citenamefont {Murray}, \citenamefont {Sloggett}, \citenamefont {Lam},
  \citenamefont {Sankrithyan}, \citenamefont {Savvides}, \citenamefont
  {Katzaros}, \citenamefont {Muller}, \citenamefont {Mitchell}, \citenamefont
  {Pollock}, \citenamefont {Lee}, \citenamefont {Dart}, \citenamefont {Barrow},
  \citenamefont {Asten}, \citenamefont {Maddever}, \citenamefont {Panjkovic},
  \citenamefont {Downey}, \citenamefont {Hoffman},\ and\ \citenamefont
  {Turner}}]{Foley1999a}%
  \BibitemOpen
  \bibfield  {author} {\bibinfo {author} {\bibfnamefont {C.}~\bibnamefont
  {Foley}}, \bibinfo {author} {\bibfnamefont {K.}~\bibnamefont {Leslie}},
  \bibinfo {author} {\bibfnamefont {R.}~\bibnamefont {Binks}}, \bibinfo
  {author} {\bibfnamefont {C.}~\bibnamefont {Lewis}}, \bibinfo {author}
  {\bibfnamefont {W.}~\bibnamefont {Murray}}, \bibinfo {author} {\bibfnamefont
  {G.}~\bibnamefont {Sloggett}}, \bibinfo {author} {\bibfnamefont
  {S.}~\bibnamefont {Lam}}, \bibinfo {author} {\bibfnamefont {B.}~\bibnamefont
  {Sankrithyan}}, \bibinfo {author} {\bibfnamefont {N.}~\bibnamefont
  {Savvides}}, \bibinfo {author} {\bibfnamefont {A.}~\bibnamefont {Katzaros}},
  \bibinfo {author} {\bibfnamefont {K.-H.}\ \bibnamefont {Muller}}, \bibinfo
  {author} {\bibfnamefont {E.}~\bibnamefont {Mitchell}}, \bibinfo {author}
  {\bibfnamefont {J.}~\bibnamefont {Pollock}}, \bibinfo {author} {\bibfnamefont
  {J.}~\bibnamefont {Lee}}, \bibinfo {author} {\bibfnamefont {D.}~\bibnamefont
  {Dart}}, \bibinfo {author} {\bibfnamefont {R.}~\bibnamefont {Barrow}},
  \bibinfo {author} {\bibfnamefont {M.}~\bibnamefont {Asten}}, \bibinfo
  {author} {\bibfnamefont {A.}~\bibnamefont {Maddever}}, \bibinfo {author}
  {\bibfnamefont {G.}~\bibnamefont {Panjkovic}}, \bibinfo {author}
  {\bibfnamefont {M.}~\bibnamefont {Downey}}, \bibinfo {author} {\bibfnamefont
  {C.}~\bibnamefont {Hoffman}}, \ and\ \bibinfo {author} {\bibfnamefont
  {R.}~\bibnamefont {Turner}},\ }\href {\doibase 10.1109/77.783852} {\bibfield
  {journal} {\bibinfo  {journal} {IEEE Transactions on Appiled
  Superconductivity}\ }\textbf {\bibinfo {volume} {9}},\ \bibinfo {pages}
  {3786} (\bibinfo {year} {1999})}\BibitemShut {NoStop}%
\bibitem [{\citenamefont {Voss}(1981)}]{Voss1981}%
  \BibitemOpen
  \bibfield  {author} {\bibinfo {author} {\bibfnamefont {R.~F.}\ \bibnamefont
  {Voss}},\ }\href {\doibase 10.1007/BF00116701} {\bibfield  {journal}
  {\bibinfo  {journal} {Journal of Low Temperature Physics}\ }\textbf {\bibinfo
  {volume} {42}},\ \bibinfo {pages} {151} (\bibinfo {year} {1981})}\BibitemShut
  {NoStop}%
\bibitem [{\citenamefont {Tesche}\ and\ \citenamefont
  {Clarke}(1977)}]{Tesche1977}%
  \BibitemOpen
  \bibfield  {author} {\bibinfo {author} {\bibfnamefont {C.~D.}\ \bibnamefont
  {Tesche}}\ and\ \bibinfo {author} {\bibfnamefont {J.}~\bibnamefont
  {Clarke}},\ }\href {\doibase 10.1007/BF00655097} {\bibfield  {journal}
  {\bibinfo  {journal} {Journal of Low Temperature Physics}\ }\textbf {\bibinfo
  {volume} {29}},\ \bibinfo {pages} {301} (\bibinfo {year} {1977})}\BibitemShut
  {NoStop}%
\bibitem [{\citenamefont {Enpuku}\ \emph {et~al.}(1993)\citenamefont {Enpuku},
  \citenamefont {Shimomura},\ and\ \citenamefont {Kisu}}]{Enpuku1993}%
  \BibitemOpen
  \bibfield  {author} {\bibinfo {author} {\bibfnamefont {K.}~\bibnamefont
  {Enpuku}}, \bibinfo {author} {\bibfnamefont {Y.}~\bibnamefont {Shimomura}}, \
  and\ \bibinfo {author} {\bibfnamefont {T.}~\bibnamefont {Kisu}},\ }\href
  {\doibase 10.1063/1.353946} {\bibfield  {journal} {\bibinfo  {journal}
  {Journal of Applied Physics}\ }\textbf {\bibinfo {volume} {73}},\ \bibinfo
  {pages} {7929} (\bibinfo {year} {1993})}\BibitemShut {NoStop}%
\bibitem [{\citenamefont {Miller}\ \emph {et~al.}(1991)\citenamefont {Miller},
  \citenamefont {Gunaratne}, \citenamefont {Huang},\ and\ \citenamefont
  {Golding}}]{Miller1991}%
  \BibitemOpen
  \bibfield  {author} {\bibinfo {author} {\bibfnamefont {J.~H.}\ \bibnamefont
  {Miller}}, \bibinfo {author} {\bibfnamefont {G.~H.}\ \bibnamefont
  {Gunaratne}}, \bibinfo {author} {\bibfnamefont {J.}~\bibnamefont {Huang}}, \
  and\ \bibinfo {author} {\bibfnamefont {T.~D.}\ \bibnamefont {Golding}},\
  }\href {\doibase 10.1063/1.105721} {\bibfield  {journal} {\bibinfo  {journal}
  {Applied Physics Letters}\ }\textbf {\bibinfo {volume} {59}},\ \bibinfo
  {pages} {3330} (\bibinfo {year} {1991})}\BibitemShut {NoStop}%
\bibitem [{\citenamefont {Gerdemann}\ \emph {et~al.}(1995)\citenamefont
  {Gerdemann}, \citenamefont {Alff}, \citenamefont {Beck}, \citenamefont
  {Froehlich}, \citenamefont {Mayer},\ and\ \citenamefont
  {Gross}}]{Gerdemann1995}%
  \BibitemOpen
  \bibfield  {author} {\bibinfo {author} {\bibfnamefont {R.}~\bibnamefont
  {Gerdemann}}, \bibinfo {author} {\bibfnamefont {L.}~\bibnamefont {Alff}},
  \bibinfo {author} {\bibfnamefont {A.}~\bibnamefont {Beck}}, \bibinfo {author}
  {\bibfnamefont {O.}~\bibnamefont {Froehlich}}, \bibinfo {author}
  {\bibfnamefont {B.}~\bibnamefont {Mayer}}, \ and\ \bibinfo {author}
  {\bibfnamefont {R.}~\bibnamefont {Gross}},\ }\href {\doibase
  10.1109/77.403295} {\bibfield  {journal} {\bibinfo  {journal} {IEEE
  Transactions on Appiled Superconductivity}\ }\textbf {\bibinfo {volume}
  {5}},\ \bibinfo {pages} {3292} (\bibinfo {year} {1995})}\BibitemShut
  {NoStop}%
\bibitem [{\citenamefont {Mitchell}\ \emph {et~al.}(2019)\citenamefont
  {Mitchell}, \citenamefont {M{\"{u}}ller}, \citenamefont {Purches},
  \citenamefont {Keenan}, \citenamefont {Lewis},\ and\ \citenamefont
  {Foley}}]{Mitchell2019}%
  \BibitemOpen
  \bibfield  {author} {\bibinfo {author} {\bibfnamefont {E.~E.}\ \bibnamefont
  {Mitchell}}, \bibinfo {author} {\bibfnamefont {K.~H.}\ \bibnamefont
  {M{\"{u}}ller}}, \bibinfo {author} {\bibfnamefont {W.~E.}\ \bibnamefont
  {Purches}}, \bibinfo {author} {\bibfnamefont {S.~T.}\ \bibnamefont {Keenan}},
  \bibinfo {author} {\bibfnamefont {C.~J.}\ \bibnamefont {Lewis}}, \ and\
  \bibinfo {author} {\bibfnamefont {C.~P.}\ \bibnamefont {Foley}},\ }\href
  {\doibase 10.1088/1361-6668/ab3fef} {\bibfield  {journal} {\bibinfo
  {journal} {Superconductor Science and Technology}\ }\textbf {\bibinfo
  {volume} {32}},\ \bibinfo {pages} {124002} (\bibinfo {year}
  {2019})}\BibitemShut {NoStop}%
\bibitem [{\citenamefont {Foglietti}\ \emph {et~al.}(1993)\citenamefont
  {Foglietti}, \citenamefont {Stawiasz}, \citenamefont {Ketchen},\ and\
  \citenamefont {Koch}}]{Foglietti1993}%
  \BibitemOpen
  \bibfield  {author} {\bibinfo {author} {\bibfnamefont {V.}~\bibnamefont
  {Foglietti}}, \bibinfo {author} {\bibfnamefont {K.}~\bibnamefont {Stawiasz}},
  \bibinfo {author} {\bibfnamefont {M.}~\bibnamefont {Ketchen}}, \ and\
  \bibinfo {author} {\bibfnamefont {R.}~\bibnamefont {Koch}},\ }\href {\doibase
  10.1109/77.251805} {\bibfield  {journal} {\bibinfo  {journal} {IEEE
  Transactions on Appiled Superconductivity}\ }\textbf {\bibinfo {volume}
  {3}},\ \bibinfo {pages} {3061} (\bibinfo {year} {1993})}\BibitemShut
  {NoStop}%
\bibitem [{\citenamefont {Krey}\ \emph {et~al.}(1999)\citenamefont {Krey},
  \citenamefont {Brugmann}, \citenamefont {Burkhardt},\ and\ \citenamefont
  {Schilling}}]{Krey1999}%
  \BibitemOpen
  \bibfield  {author} {\bibinfo {author} {\bibfnamefont {S.}~\bibnamefont
  {Krey}}, \bibinfo {author} {\bibfnamefont {O.}~\bibnamefont {Brugmann}},
  \bibinfo {author} {\bibfnamefont {H.}~\bibnamefont {Burkhardt}}, \ and\
  \bibinfo {author} {\bibfnamefont {M.}~\bibnamefont {Schilling}},\ }\href
  {\doibase 10.1109/77.783759} {\bibfield  {journal} {\bibinfo  {journal} {IEEE
  Transactions on Appiled Superconductivity}\ }\textbf {\bibinfo {volume}
  {9}},\ \bibinfo {pages} {3401} (\bibinfo {year} {1999})}\BibitemShut
  {NoStop}%
\bibitem [{\citenamefont {Berggren}(2012)}]{Berggren2012}%
  \BibitemOpen
  \bibfield  {author} {\bibinfo {author} {\bibfnamefont {S.}~\bibnamefont
  {Berggren}},\ }\emph {\bibinfo {title} {{Computational and Mathematical
  Modeling of Coupled Superconducting Quantum Interference Devices}}},\
  \href@noop {} {Ph.D. thesis},\ \bibinfo  {school} {Claremont Graduate
  University; San Diego State University} (\bibinfo {year} {2012})\BibitemShut
  {NoStop}%
\bibitem [{\citenamefont {Berggren}\ and\ \citenamefont {{De
  Escobar}}(2015)}]{Berggren2015}%
  \BibitemOpen
  \bibfield  {author} {\bibinfo {author} {\bibfnamefont {S.}~\bibnamefont
  {Berggren}}\ and\ \bibinfo {author} {\bibfnamefont {A.~L.}\ \bibnamefont {{De
  Escobar}}},\ }\href {\doibase 10.1109/TASC.2014.2359375} {\bibfield
  {journal} {\bibinfo  {journal} {IEEE Transactions on Applied
  Superconductivity}\ }\textbf {\bibinfo {volume} {25}},\ \bibinfo {pages} {1}
  (\bibinfo {year} {2015})}\BibitemShut {NoStop}%
\bibitem [{\citenamefont {M{\"{u}}ller}\ and\ \citenamefont
  {Mitchell}(2021)}]{Muller2021}%
  \BibitemOpen
  \bibfield  {author} {\bibinfo {author} {\bibfnamefont {K.-H.}\ \bibnamefont
  {M{\"{u}}ller}}\ and\ \bibinfo {author} {\bibfnamefont {E.~E.}\ \bibnamefont
  {Mitchell}},\ }\href {\doibase 10.1103/PhysRevB.103.054509} {\bibfield
  {journal} {\bibinfo  {journal} {Physical Review B}\ }\textbf {\bibinfo
  {volume} {103}},\ \bibinfo {pages} {054509} (\bibinfo {year}
  {2021})}\BibitemShut {NoStop}%
\bibitem [{\citenamefont {Oppenl{\"{a}}nder}\ \emph {et~al.}(2000)\citenamefont
  {Oppenl{\"{a}}nder}, \citenamefont {H{\"{a}}ussler},\ and\ \citenamefont
  {Schopohl}}]{Oppenlander2000}%
  \BibitemOpen
  \bibfield  {author} {\bibinfo {author} {\bibfnamefont {J.}~\bibnamefont
  {Oppenl{\"{a}}nder}}, \bibinfo {author} {\bibfnamefont {C.}~\bibnamefont
  {H{\"{a}}ussler}}, \ and\ \bibinfo {author} {\bibfnamefont {N.}~\bibnamefont
  {Schopohl}},\ }\href {\doibase 10.1103/PhysRevB.63.024511} {\bibfield
  {journal} {\bibinfo  {journal} {Physical Review B}\ }\textbf {\bibinfo
  {volume} {63}},\ \bibinfo {pages} {024511} (\bibinfo {year}
  {2000})}\BibitemShut {NoStop}%
\bibitem [{\citenamefont {Caputo}\ \emph {et~al.}(2005)\citenamefont {Caputo},
  \citenamefont {Tomes}, \citenamefont {Oppenl{\"{a}}nder}, \citenamefont
  {H{\"{a}}ussler}, \citenamefont {Friesch}, \citenamefont {Tr{\"{a}}uble},\
  and\ \citenamefont {Schopohl}}]{Caputo2005}%
  \BibitemOpen
  \bibfield  {author} {\bibinfo {author} {\bibfnamefont {P.}~\bibnamefont
  {Caputo}}, \bibinfo {author} {\bibfnamefont {J.}~\bibnamefont {Tomes}},
  \bibinfo {author} {\bibfnamefont {J.}~\bibnamefont {Oppenl{\"{a}}nder}},
  \bibinfo {author} {\bibfnamefont {C.}~\bibnamefont {H{\"{a}}ussler}},
  \bibinfo {author} {\bibfnamefont {A.}~\bibnamefont {Friesch}}, \bibinfo
  {author} {\bibfnamefont {T.}~\bibnamefont {Tr{\"{a}}uble}}, \ and\ \bibinfo
  {author} {\bibfnamefont {N.}~\bibnamefont {Schopohl}},\ }\href {\doibase
  10.1109/TASC.2003.850193} {\bibfield  {journal} {\bibinfo  {journal} {IEEE
  Transactions on Applied Superconductivity}\ }\textbf {\bibinfo {volume}
  {15}},\ \bibinfo {pages} {1044} (\bibinfo {year} {2005})}\BibitemShut
  {NoStop}%
\bibitem [{\citenamefont {H{\"{a}}ussler}\ \emph {et~al.}(2001)\citenamefont
  {H{\"{a}}ussler}, \citenamefont {Oppenl{\"{a}}nder},\ and\ \citenamefont
  {Schopohl}}]{Haussler2001}%
  \BibitemOpen
  \bibfield  {author} {\bibinfo {author} {\bibfnamefont {C.}~\bibnamefont
  {H{\"{a}}ussler}}, \bibinfo {author} {\bibfnamefont {J.}~\bibnamefont
  {Oppenl{\"{a}}nder}}, \ and\ \bibinfo {author} {\bibfnamefont
  {N.}~\bibnamefont {Schopohl}},\ }\href {\doibase 10.1063/1.1334374}
  {\bibfield  {journal} {\bibinfo  {journal} {Journal of Applied Physics}\
  }\textbf {\bibinfo {volume} {89}},\ \bibinfo {pages} {1875} (\bibinfo {year}
  {2001})}\BibitemShut {NoStop}%
\bibitem [{\citenamefont {Oppenl{\"{a}}nder}\ \emph {et~al.}(2002)\citenamefont
  {Oppenl{\"{a}}nder}, \citenamefont {H{\"{a}}ussler}, \citenamefont
  {Tr{\"{a}}uble},\ and\ \citenamefont {Schopohl}}]{Oppenlander2002}%
  \BibitemOpen
  \bibfield  {author} {\bibinfo {author} {\bibfnamefont {J.}~\bibnamefont
  {Oppenl{\"{a}}nder}}, \bibinfo {author} {\bibfnamefont {C.}~\bibnamefont
  {H{\"{a}}ussler}}, \bibinfo {author} {\bibfnamefont {T.}~\bibnamefont
  {Tr{\"{a}}uble}}, \ and\ \bibinfo {author} {\bibfnamefont {N.}~\bibnamefont
  {Schopohl}},\ }\href {\doibase 10.1016/S0921-4534(01)01151-0} {\bibfield
  {journal} {\bibinfo  {journal} {Physica C: Superconductivity}\ }\textbf
  {\bibinfo {volume} {368}},\ \bibinfo {pages} {119} (\bibinfo {year}
  {2002})}\BibitemShut {NoStop}%
\bibitem [{\citenamefont {Longhini}\ \emph {et~al.}(2011)\citenamefont
  {Longhini}, \citenamefont {Berggren}, \citenamefont {Palacios}, \citenamefont
  {In},\ and\ \citenamefont {{De Escobar}}}]{Longhini2011}%
  \BibitemOpen
  \bibfield  {author} {\bibinfo {author} {\bibfnamefont {P.}~\bibnamefont
  {Longhini}}, \bibinfo {author} {\bibfnamefont {S.}~\bibnamefont {Berggren}},
  \bibinfo {author} {\bibfnamefont {A.}~\bibnamefont {Palacios}}, \bibinfo
  {author} {\bibfnamefont {V.}~\bibnamefont {In}}, \ and\ \bibinfo {author}
  {\bibfnamefont {A.~L.}\ \bibnamefont {{De Escobar}}},\ }\href {\doibase
  10.1109/TASC.2011.2105455} {\bibfield  {journal} {\bibinfo  {journal} {IEEE
  Transactions on Applied Superconductivity}\ }\textbf {\bibinfo {volume}
  {21}},\ \bibinfo {pages} {391} (\bibinfo {year} {2011})}\BibitemShut
  {NoStop}%
\bibitem [{\citenamefont {Schultze}\ \emph {et~al.}(2006)\citenamefont
  {Schultze}, \citenamefont {IJsselsteijn},\ and\ \citenamefont
  {Meyer}}]{Schultze2006}%
  \BibitemOpen
  \bibfield  {author} {\bibinfo {author} {\bibfnamefont {V.}~\bibnamefont
  {Schultze}}, \bibinfo {author} {\bibfnamefont {R.}~\bibnamefont
  {IJsselsteijn}}, \ and\ \bibinfo {author} {\bibfnamefont {H.-G.}\
  \bibnamefont {Meyer}},\ }\href {\doibase 10.1088/0953-2048/19/5/S52}
  {\bibfield  {journal} {\bibinfo  {journal} {Superconductor Science and
  Technology}\ }\textbf {\bibinfo {volume} {19}},\ \bibinfo {pages} {S411}
  (\bibinfo {year} {2006})}\BibitemShut {NoStop}%
\bibitem [{\citenamefont {Kornev}\ \emph
  {et~al.}(2009{\natexlab{a}})\citenamefont {Kornev}, \citenamefont {Soloviev},
  \citenamefont {Klenov}, \citenamefont {Filippov}, \citenamefont {Engseth},\
  and\ \citenamefont {Mukhanov}}]{Kornev2009b}%
  \BibitemOpen
  \bibfield  {author} {\bibinfo {author} {\bibfnamefont {V.}~\bibnamefont
  {Kornev}}, \bibinfo {author} {\bibfnamefont {I.}~\bibnamefont {Soloviev}},
  \bibinfo {author} {\bibfnamefont {N.}~\bibnamefont {Klenov}}, \bibinfo
  {author} {\bibfnamefont {T.}~\bibnamefont {Filippov}}, \bibinfo {author}
  {\bibfnamefont {H.}~\bibnamefont {Engseth}}, \ and\ \bibinfo {author}
  {\bibfnamefont {O.}~\bibnamefont {Mukhanov}},\ }\href {\doibase
  10.1109/TASC.2009.2019589} {\bibfield  {journal} {\bibinfo  {journal} {IEEE
  Transactions on Applied Superconductivity}\ }\textbf {\bibinfo {volume}
  {19}},\ \bibinfo {pages} {916} (\bibinfo {year}
  {2009}{\natexlab{a}})}\BibitemShut {NoStop}%
\bibitem [{\citenamefont {Kornev}\ \emph {et~al.}(2011)\citenamefont {Kornev},
  \citenamefont {Soloviev}, \citenamefont {Klenov},\ and\ \citenamefont
  {Mukhanov}}]{Kornev2011}%
  \BibitemOpen
  \bibfield  {author} {\bibinfo {author} {\bibfnamefont {V.~K.}\ \bibnamefont
  {Kornev}}, \bibinfo {author} {\bibfnamefont {I.~I.}\ \bibnamefont
  {Soloviev}}, \bibinfo {author} {\bibfnamefont {N.~V.}\ \bibnamefont
  {Klenov}}, \ and\ \bibinfo {author} {\bibfnamefont {O.~A.}\ \bibnamefont
  {Mukhanov}},\ }\href {\doibase 10.1109/TASC.2010.2095451} {\bibfield
  {journal} {\bibinfo  {journal} {IEEE Transactions on Applied
  Superconductivity}\ }\textbf {\bibinfo {volume} {21}},\ \bibinfo {pages}
  {394} (\bibinfo {year} {2011})}\BibitemShut {NoStop}%
\bibitem [{\citenamefont {Mitchell}\ \emph {et~al.}(2016)\citenamefont
  {Mitchell}, \citenamefont {Hannam}, \citenamefont {Lazar}, \citenamefont
  {Leslie}, \citenamefont {Lewis}, \citenamefont {Grancea}, \citenamefont
  {Keenan}, \citenamefont {Lam},\ and\ \citenamefont {Foley}}]{Mitchell2016}%
  \BibitemOpen
  \bibfield  {author} {\bibinfo {author} {\bibfnamefont {E.~E.}\ \bibnamefont
  {Mitchell}}, \bibinfo {author} {\bibfnamefont {K.~E.}\ \bibnamefont
  {Hannam}}, \bibinfo {author} {\bibfnamefont {J.}~\bibnamefont {Lazar}},
  \bibinfo {author} {\bibfnamefont {K.~E.}\ \bibnamefont {Leslie}}, \bibinfo
  {author} {\bibfnamefont {C.~J.}\ \bibnamefont {Lewis}}, \bibinfo {author}
  {\bibfnamefont {A.}~\bibnamefont {Grancea}}, \bibinfo {author} {\bibfnamefont
  {S.~T.}\ \bibnamefont {Keenan}}, \bibinfo {author} {\bibfnamefont {S.~K.}\
  \bibnamefont {Lam}}, \ and\ \bibinfo {author} {\bibfnamefont {C.~P.}\
  \bibnamefont {Foley}},\ }\href {\doibase 10.1088/0953-2048/29/6/06LT01}
  {\bibfield  {journal} {\bibinfo  {journal} {Superconductor Science and
  Technology}\ }\textbf {\bibinfo {volume} {29}},\ \bibinfo {pages} {1}
  (\bibinfo {year} {2016})}\BibitemShut {NoStop}%
\bibitem [{\citenamefont {Taylor}\ \emph {et~al.}(2016)\citenamefont {Taylor},
  \citenamefont {Berggren}, \citenamefont {O'Brien}, \citenamefont {DeAndrade},
  \citenamefont {Higa},\ and\ \citenamefont {de~Escobar}}]{Taylor2016}%
  \BibitemOpen
  \bibfield  {author} {\bibinfo {author} {\bibfnamefont {B.~J.}\ \bibnamefont
  {Taylor}}, \bibinfo {author} {\bibfnamefont {S.~A.~E.}\ \bibnamefont
  {Berggren}}, \bibinfo {author} {\bibfnamefont {M.~C.}\ \bibnamefont
  {O'Brien}}, \bibinfo {author} {\bibfnamefont {M.~C.}\ \bibnamefont
  {DeAndrade}}, \bibinfo {author} {\bibfnamefont {B.~A.}\ \bibnamefont {Higa}},
  \ and\ \bibinfo {author} {\bibfnamefont {A.~M.~L.}\ \bibnamefont
  {de~Escobar}},\ }\href {\doibase 10.1088/0953-2048/29/8/084003} {\bibfield
  {journal} {\bibinfo  {journal} {Superconductor Science and Technology}\
  }\textbf {\bibinfo {volume} {29}},\ \bibinfo {pages} {084003} (\bibinfo
  {year} {2016})}\BibitemShut {NoStop}%
\bibitem [{\citenamefont {Cybart}\ \emph {et~al.}(2012)\citenamefont {Cybart},
  \citenamefont {Dalichaouch}, \citenamefont {Wu}, \citenamefont {Anton},
  \citenamefont {Drisko}, \citenamefont {Parker}, \citenamefont {Harteneck},\
  and\ \citenamefont {Dynes}}]{Cybart2012}%
  \BibitemOpen
  \bibfield  {author} {\bibinfo {author} {\bibfnamefont {S.~A.}\ \bibnamefont
  {Cybart}}, \bibinfo {author} {\bibfnamefont {T.~N.}\ \bibnamefont
  {Dalichaouch}}, \bibinfo {author} {\bibfnamefont {S.~M.}\ \bibnamefont {Wu}},
  \bibinfo {author} {\bibfnamefont {S.~M.}\ \bibnamefont {Anton}}, \bibinfo
  {author} {\bibfnamefont {J.~A.}\ \bibnamefont {Drisko}}, \bibinfo {author}
  {\bibfnamefont {J.~M.}\ \bibnamefont {Parker}}, \bibinfo {author}
  {\bibfnamefont {B.~D.}\ \bibnamefont {Harteneck}}, \ and\ \bibinfo {author}
  {\bibfnamefont {R.~C.}\ \bibnamefont {Dynes}},\ }\href {\doibase
  10.1063/1.4754422} {\bibfield  {journal} {\bibinfo  {journal} {Journal of
  Applied Physics}\ }\textbf {\bibinfo {volume} {112}},\ \bibinfo {pages}
  {063911} (\bibinfo {year} {2012})}\BibitemShut {NoStop}%
\bibitem [{\citenamefont {Dalichaouch}\ \emph {et~al.}(2014)\citenamefont
  {Dalichaouch}, \citenamefont {Cybart},\ and\ \citenamefont
  {Dynes}}]{Dalichaouch2014}%
  \BibitemOpen
  \bibfield  {author} {\bibinfo {author} {\bibfnamefont {T.~N.}\ \bibnamefont
  {Dalichaouch}}, \bibinfo {author} {\bibfnamefont {S.~A.}\ \bibnamefont
  {Cybart}}, \ and\ \bibinfo {author} {\bibfnamefont {R.~C.}\ \bibnamefont
  {Dynes}},\ }\href {\doibase 10.1088/0953-2048/27/6/065006} {\bibfield
  {journal} {\bibinfo  {journal} {Superconductor Science and Technology}\
  }\textbf {\bibinfo {volume} {27}},\ \bibinfo {pages} {065006} (\bibinfo
  {year} {2014})}\BibitemShut {NoStop}%
\bibitem [{\citenamefont {Newrock}\ \emph {et~al.}(2000)\citenamefont
  {Newrock}, \citenamefont {Lobb}, \citenamefont {Geigenm{\"{u}}ller},\ and\
  \citenamefont {Octavio}}]{Newrock2000}%
  \BibitemOpen
  \bibfield  {author} {\bibinfo {author} {\bibfnamefont {R.}~\bibnamefont
  {Newrock}}, \bibinfo {author} {\bibfnamefont {C.}~\bibnamefont {Lobb}},
  \bibinfo {author} {\bibfnamefont {U.}~\bibnamefont {Geigenm{\"{u}}ller}}, \
  and\ \bibinfo {author} {\bibfnamefont {M.}~\bibnamefont {Octavio}},\ }in\
  \href {\doibase 10.1016/S0081-1947(08)60250-7} {\emph {\bibinfo {booktitle}
  {Solid State Physics - Advances in Research and Applications}}},\
  Vol.~\bibinfo {volume} {54}\ (\bibinfo {year} {2000})\ pp.\ \bibinfo {pages}
  {263--512}\BibitemShut {NoStop}%
\bibitem [{\citenamefont {Tinkham}(2004)}]{Tinkham2004}%
  \BibitemOpen
  \bibfield  {author} {\bibinfo {author} {\bibfnamefont {M.}~\bibnamefont
  {Tinkham}},\ }\href {http://www.worldcat.org/isbn/0486435032} {\emph
  {\bibinfo {title} {{Introduction to Superconductivity}}}},\ \bibinfo
  {edition} {2nd}\ ed.\ (\bibinfo  {publisher} {Dover Publications},\ \bibinfo
  {address} {New York},\ \bibinfo {year} {2004})\BibitemShut {NoStop}%
\bibitem [{\citenamefont {London}\ and\ \citenamefont
  {London}(1935)}]{London1935}%
  \BibitemOpen
  \bibfield  {author} {\bibinfo {author} {\bibfnamefont {F.}~\bibnamefont
  {London}}\ and\ \bibinfo {author} {\bibfnamefont {H.}~\bibnamefont
  {London}},\ }\href {\doibase 10.1098/rspa.1935.0048} {\bibfield  {journal}
  {\bibinfo  {journal} {Proceedings of the Royal Society of London. Series A -
  Mathematical and Physical Sciences}\ }\textbf {\bibinfo {volume} {149}},\
  \bibinfo {pages} {71} (\bibinfo {year} {1935})}\BibitemShut {NoStop}%
\bibitem [{\citenamefont {Pearl}(1964)}]{Pearl1964}%
  \BibitemOpen
  \bibfield  {author} {\bibinfo {author} {\bibfnamefont {J.}~\bibnamefont
  {Pearl}},\ }\href {\doibase 10.1063/1.1754056} {\bibfield  {journal}
  {\bibinfo  {journal} {Applied Physics Letters}\ }\textbf {\bibinfo {volume}
  {5}},\ \bibinfo {pages} {65} (\bibinfo {year} {1964})}\BibitemShut {NoStop}%
\bibitem [{\citenamefont {Hoer}\ and\ \citenamefont {Love}(1965)}]{Hoer1965}%
  \BibitemOpen
  \bibfield  {author} {\bibinfo {author} {\bibfnamefont {C.}~\bibnamefont
  {Hoer}}\ and\ \bibinfo {author} {\bibfnamefont {C.}~\bibnamefont {Love}},\
  }\href {\doibase 10.6028/jres.069C.016} {\bibfield  {journal} {\bibinfo
  {journal} {Journal of Research of the National Bureau of Standards, Section
  C: Engineering and Instrumentation}\ }\textbf {\bibinfo {volume} {69C}},\
  \bibinfo {pages} {127} (\bibinfo {year} {1965})}\BibitemShut {NoStop}%
\bibitem [{\citenamefont {Khapaev}\ \emph {et~al.}(2001)\citenamefont
  {Khapaev}, \citenamefont {Kidiyarova-Shevchenko}, \citenamefont {Magnelind},\
  and\ \citenamefont {Kupriyanov}}]{Khapaev2001}%
  \BibitemOpen
  \bibfield  {author} {\bibinfo {author} {\bibfnamefont {M.}~\bibnamefont
  {Khapaev}}, \bibinfo {author} {\bibfnamefont {A.}~\bibnamefont
  {Kidiyarova-Shevchenko}}, \bibinfo {author} {\bibfnamefont {P.}~\bibnamefont
  {Magnelind}}, \ and\ \bibinfo {author} {\bibfnamefont {M.}~\bibnamefont
  {Kupriyanov}},\ }\href {\doibase 10.1109/77.919537} {\bibfield  {journal}
  {\bibinfo  {journal} {IEEE Transactions on Appiled Superconductivity}\
  }\textbf {\bibinfo {volume} {11}},\ \bibinfo {pages} {1090} (\bibinfo {year}
  {2001})}\BibitemShut {NoStop}%
\bibitem [{\citenamefont {Kamon}\ \emph {et~al.}(1994)\citenamefont {Kamon},
  \citenamefont {Ttsuk},\ and\ \citenamefont {White}}]{Kamon1994}%
  \BibitemOpen
  \bibfield  {author} {\bibinfo {author} {\bibfnamefont {M.}~\bibnamefont
  {Kamon}}, \bibinfo {author} {\bibfnamefont {M.}~\bibnamefont {Ttsuk}}, \ and\
  \bibinfo {author} {\bibfnamefont {J.}~\bibnamefont {White}},\ }\href
  {\doibase 10.1109/22.310584} {\bibfield  {journal} {\bibinfo  {journal} {IEEE
  Transactions on Microwave Theory and Techniques}\ }\textbf {\bibinfo {volume}
  {42}},\ \bibinfo {pages} {1750} (\bibinfo {year} {1994})}\BibitemShut
  {NoStop}%
\bibitem [{\citenamefont {Tausch}\ and\ \citenamefont
  {White}(1999)}]{Tausch1999}%
  \BibitemOpen
  \bibfield  {author} {\bibinfo {author} {\bibfnamefont {J.}~\bibnamefont
  {Tausch}}\ and\ \bibinfo {author} {\bibfnamefont {J.}~\bibnamefont {White}},\
  }\href {\doibase 10.1109/22.740070} {\bibfield  {journal} {\bibinfo
  {journal} {IEEE Transactions on Microwave Theory and Techniques}\ }\textbf
  {\bibinfo {volume} {47}},\ \bibinfo {pages} {18} (\bibinfo {year}
  {1999})}\BibitemShut {NoStop}%
\bibitem [{\citenamefont {Kornev}\ \emph
  {et~al.}(2009{\natexlab{b}})\citenamefont {Kornev}, \citenamefont {Soloviev},
  \citenamefont {Klenov},\ and\ \citenamefont {Mukhanov}}]{Kornev2009a}%
  \BibitemOpen
  \bibfield  {author} {\bibinfo {author} {\bibfnamefont {V.}~\bibnamefont
  {Kornev}}, \bibinfo {author} {\bibfnamefont {I.}~\bibnamefont {Soloviev}},
  \bibinfo {author} {\bibfnamefont {N.}~\bibnamefont {Klenov}}, \ and\ \bibinfo
  {author} {\bibfnamefont {O.}~\bibnamefont {Mukhanov}},\ }\href {\doibase
  10.1109/TASC.2009.2019543} {\bibfield  {journal} {\bibinfo  {journal} {IEEE
  Transactions on Applied Superconductivity}\ }\textbf {\bibinfo {volume}
  {19}},\ \bibinfo {pages} {741} (\bibinfo {year}
  {2009}{\natexlab{b}})}\BibitemShut {NoStop}%
\end{thebibliography}%

\end{document}